\documentclass[12pt,preprint]{aastex}

\usepackage{epsfig}
\usepackage{natbib}

\shorttitle{ANNIHILATION EMISSION} \shortauthors{CHENG ET AL.}

\begin{document}


\title{ANNIHILATION EMISSION FROM THE GALACTIC BLACK HOLE}

\author{K. S. Cheng\altaffilmark{1},  D. O. Chernyshov\altaffilmark{1,2}, and V. A. Dogiel\altaffilmark{1,3}}

\altaffiltext{1}{
 Department of Physics, University of Hong Kong, Pokfulam Road, Hong Kong, China;}
 \altaffiltext{2}{ Moscow Institute of Physics and Technology, Institutskii lane, 141700 Moscow
Region,
  Dolgoprudnii, Russia;}
 \altaffiltext{3}{I.E.Tamm Theoretical Physics Division of P.N.Lebedev Institute, Leninskii pr, 53,
119991 Moscow, Russia.}


\begin{abstract}
Both diffuse high energy gamma-rays and an extended electron-positron
annihilation line emission have been observed in the Galactic Center (GC)
region. Although X-ray observations indicate that the galactic black hole Sgr
A$^*$ is inactive now, we suggest that Sgr A$^*$ can become active when a captured star is
tidally disrupted and matter is accreted into the black hole.  As a consequence
the galactic black hole could be a powerful source of relativistic protons. 
We are able to explain the current observed diffuse gamma-rays and the very detailed
511 keV annihilation line of secondary positrons by $p-p$ collisions 
of such protons, with appropriate injection times and energy. 
Relativistic protons could have been injected into the ambient material if the black hole captured    
a 50M$_\odot$ star at several tens million years ago. 
An alternative possibility is that the black hole 
continues to capture stars with $\sim$1M$_\odot$ every hundred thousand years. Secondary
positrons produced by $p-p$ collisions at energies $\ga 30$ MeV are cooled down to
thermal energies by Coulomb collisions, and annihilate in the warm neutral and
ionized phases of the interstellar medium with temperatures about several eV,
because the annihilation cross-section reaches its maximum at these
temperatures. It takes about ten million years for the positrons to cool down to thermal temperatures so they can 
diffuse into a very large extended region around the Galactic center. A much
more recent star capture may be also able to account for recent TeV
observations within 10 pc of the galactic center as well as for the
unidentified GeV gamma-ray sources found by EGRET at GC. The spectral
difference between the GeV flux and the TeV flux could be explained naturally in this
model as well.
\end{abstract}


\keywords{comic rays : general - Galaxy : center - Galaxies : gamma-rays - black hole - radiation mechanisms : 
nonthermal}


\section{Introduction}
The annihilation of cosmic ray positrons produced as a result of cosmic ray
proton collisions with the ambient plasma was discussed in a number of papers
\citep[see, e.g.,][ etc.]{ginz64,haya64,steck,buss}. A typical cosmic ray
positron resulting from secondary $\pi$-meson decay may undergo one of three fates: 1)
escape from the Galaxy, 2) annihilate with an electron while at relativistic
energy ({\it in-flight annihilation}), or 3) lose almost all its energy before
annihilation. One can consider that annihilation may occur either between free
electrons and positrons ({\it free annihilation}), or through the formation of
the intermediate bound state of positronium. Positronium can form in either a
singlet state, which annihilate into two 511 keV photons, or in a triplet state
which decays by three photon annihilation producing a continuum emission
below 511 keV  \citep[see][]{ore}.

The cross-section for positron annihilation as a function of energy was given
first by \citet{dir} \citep[see also][]{heit}.

Radiation from electron-positron annihilation from the central region of our
Galaxy was first reported more than 30 years ago \citep{joh}. In 1978 it was
reported that there is a flux of the positron annihilation line
\citep[see][]{lev} which might be produced in a compact source at or near the
Galactic center.  First estimations of the morphology of this line was obtained
by the OSSE telescope on-board the Compton observatory  \citep[see,
e.g.,][]{cheng97}. With the launch of the INTEGRAL telescope high resolution
spectroscopy with reasonable angular resolutions became available. These
measurements showed a flux of the annihilation emission from the central bulge.
The bulge annihilation emission is highly symmetric with an extension of $8^o$.
The INTEGRAL team measured a 511 keV line, of width 2.2 keV, with a flux
 $(1.01\pm 0.02)\times 10^{-3}$ ph cm$^{-2}$s$^{-1}$ and
an ortho-positronium annihilation continuum flux equal TO $(4.3\pm 0.3)\times 10^{-3}$ ph
cm$^{-2}$s$^{-1}$ \citep[see][]{teer, knoed,chur_INTEGRAL,jean}.

 From the ratio between  line
 and continuum intensities near and below 511 keV it was concluded that positrons annihilate
 via positronium formation. The fraction of positronium was estimated to be in the region of
 92-97\% \citep{weid,jean}. From  the observed width of the annihilation line
 it was concluded  \citep[see][]{chur_INTEGRAL,jean} that the annihilation of thermal positrons takes place
in a relatively warm (several eV) and lowly ionized ($\sim 0.1$) interstellar
 medium.

 The origin of such positrons is still poorly understood. A large variety of positron
 sources have been proposed. Among them novae and supernovae stars
 \citep{chan,dermm}. However, as follows from
\citet{weid05}  the annihilation of positrons appears to be even more
concentrated in the bulge than are old stellar populations such as Type Ia
supernovae, novae, or low-mass X-ray binaries. New, speculative, physics such
as positron production from hypernovae stars \citep{casse}, light dark matter
\citep[see][]{boehm}  etc. has begun to be discussed as a possible solution. A
supermassive black hole of $2.45\times 10^6$ $M_\odot$  is another candidate as
a  source of positrons.  Thus, \citet{zhel} concluded that the annihilation
emission is consistent with a model in which electron-positron pairs, created
in the vicinity of a black hole, are trapped in the magnetosphere supported by
an accretion disk.

Secondary positrons generated as a result of cosmic ray collisions with
the ambient plasma is another alternative model. \citet{melia,fatuz} and
\citet{fat} suggested a supernova coinciding with the central radio source Sgr
A East as an emitter of protons which can  produce gamma-rays as well as high
energy secondary positrons. However, they showed from simplified equations that
Sgr A East could not be the source of annihilation radiation from the Galactic
center because, based on their estimates, the thermalization time was too
long. Besides, the annihilation emission appears to be diffuse. So far, there
is no evidence for emission from point-like sources \citep{knoed,bouch}.
 On the other hand, the EGRET telescope found a flux of gamma-rays
 from the Galactic center  at energies $>500$ MeV \citep{mayer} which
 was seen as a strong excess of gamma-rays
peaking  in an error circle of $0.2^o$ radius surrounded by a  strong emission
maximum within $\sim 5-8^o$ degrees from the Galactic center.

In our model we assume the black hole is a source of high energy protons
generated by star accretion. They produce secondary  gamma-rays and
relativistic positrons as a result of $p-p$ collisions. The secondary positrons requires
over 10 million years to cool down before annihilation. Hence they are able to
propagate far away from the central source during their lifetime and to fill a
sphere with the radius about several hundred pc.  Therefore the annihilation
emission can be seen as diffuse if these protons were ejected a relatively long
time ago. From kinetic equations we shall show that processes of Coulomb
collisions are effective enough to cool down these relativistic positrons and
to thermalize them before their annihilation, which can explain the origin of the
annihilation emission from the Galactic center.

\section{Energy Release Supplied by a Black Hole}

We will first summarize how much energy must be released from the Galactic black hole in order to explain the 
observed X-rays, diffuse gamma-rays, the annihilation emission and TeV emission.

\subsection{ X-rays}
 The ASCA telescope found  hard X-ray emission from in the energy range 2-10 keV
 the
central part of the Galactic plane 
{\citep{koyama}. The total thermal energy of the hot plasma component inside
a $1^0$ region around the Galactic center was estimated to be $10^{52}-10^{53}$ erg.
A large energy generation rate in the Galactic Center is required $\sim
10^{41}-10^{42}$ erg s$^{-1}$ and a supernova origin of this emission
seems to be implausible.

Recently, {\it Chrandra} observations  showed intense X-ray emission at
energy $E_x\sim 8$ keV from the inner 20 pc of the Galaxy, with a
luminosity $L_x\sim 10^{38}$ erg s$^{-1}$. We know of no class of object which
can provide this luminosity there. Therefore this emission was interpreted as 
thermal emission of an optically thin plasma with a temperature of 8 keV. However,
for the gravitational potential near the Galactic center \citep[see][]{breit}
the escape velocity is about 900 km s$^{-1}$. The sound speed of an 8 keV plasma
is 1500 km s$^{-1}$ and the required power necessary to compensate the plasma
outflow is $\sim 10^{40}$ erg s$^{-1}$. This is equivalent to the entire of
supernova kinetic energy output occurring every 3,000 years, and is unreasonably
high \citep{muno04}.

\subsection{ Gamma-rays.}
The EGRET telescope (see \citet{mayer}) found a gamma-ray flux toward the Galactic Center
of the order of $2\times 10^{37}$ erg s$^{-1}$ for
 energies $E_\gamma>500$ MeV in an error circle of $0.2^o$
radius.  The photon spectrum from the center (ph cm$^{-2}$s$^{-1}$)
can be represented as

\begin{eqnarray}
&&F(E_\gamma)\simeq 2.2\times 10^{-10}\left({E_\gamma\over
{1900~MeV}}\right)^{-1.3}\\
&&\mbox{for}~E_\gamma< 1900~\mbox{MeV}\nonumber
\end{eqnarray}
and
\begin{eqnarray}
 &&F(E_\gamma)\simeq 2.2\times 10^{-10} \left({E_\gamma\over
{1900~MeV}}\right)^{-3.1}\label{fgamma} \\
&&\mbox{for}~E_\gamma> 1900~\mbox{MeV}\nonumber
\end{eqnarray}

The bulk of the emission is most compatible with it originating in a volume
with a radius 85 pc though, in view of uncertainties, the source excess is
marginally compatible with emission from a single compact object.  If we assume
that the central gamma-ray source is due to injection of relativistic protons we
can estimate their energy. The flux of gamma-rays from the central source is
about $F_\gamma\simeq 2\cdot 10^{37}$ erg s$^{-1}$. The lifetime of
relativistic protons in the interstellar medium with a density $n\simeq 1-10$
cm$^{-3}$ is about $\tau_p\simeq 10^{14}-10^{15}$s, which gives a total energy
in relativistic protons $W_p\simeq F_\gamma\cdot \tau_p\sim 10^{52}$ erg. This
is more than can be supplied by the most powerful known galactic source of
energy, such as a supernova, whose total energy release is about $10^{51}$ erg.

\subsection{Annihilation line emission}

On the basis of the reported flux of annihilation line emission  \citep[see][]{teer,
knoed,chur_INTEGRAL,jean}, the annihilation rate of positrons is roughly $\sim
10^{43}$s$^{-1}$. Assuming that each interaction of a relativistic proton
with an average energy about 1 GeV produces one secondary positron, we find
that the injected energy rate {of protons is $\dot{W}_p \sim 10^{40}$erg
s$^{-1}$. The total amount of injected proton energy must be of the order
(or smaller than) $W_p \sim \dot{W}_p \tau_c \sim 10^{54}$erg, where $\tau_c
\sim 10^{14}$s is the cooling time of relativistic positrons in the
interstellar medium with a gas density $n\sim 1$ cm$^{-3}$. Such a huge amount
of energy is the most important constraint of the model. In section 2.5, we
will discuss some possible energy injection mechanisms.

\subsection{ TeV gamma-rays}

TeV gamma-ray emission from the Galactic black hole Sgr A* was first
reported by the Whipple group (Buckley et al. 1997). More recently it
has been confirmed by three independent groups, Whipple \citep{kosa}, CANGAROO
\citep{tsu}, and HESS \citep{aha04} at a luminosity $>10^{35}$ ergs s$^{-1}$.
If these TeV gamma-rays are produced via $p$-$p$ collisions, the possible input
proton energy could be $10^{50-52} erg$ depending on the proton injection
spectrum and the diffusion coefficient (Aharonian and Neronov 2005a).

Thus, observational data indicate a huge energy release at the Galactic
center with an energy of the order $>10^{52}$ erg, which cannot be supplied by
known Galactic sources, and a significant part of this energy is enclosed in the
form of relativistic protons. One can assume that such energy can be released
when a star is disrupted near a black hole.

\subsection{ Theoretical estimates.}
The rate at which a massive black hole in a dense star cluster tidally 
disrupts and swallows stars has been studied extensively 
\citep[e.g.][]{hills,bah,light}. Basically when a star trajectory happens to be sufficiently close to a
massive black hole, the star would be captured and eventually disrupted by tidal forces. After a dynamical
time-scale (orbital time-scale), the debris of a tidally disrupted star will form a transient accretion
disk around the massive black hole, with a radius typically comparable to the tidal capture radius \citep{rees}. Rees 
has also argued that most of the debris material will be swallowed by a black hole with a mass
$\sim 10^6 M_{\odot}$ on a time scale of $\sim 1$ yr for a thick hot ring, or $\sim 10^2$ yrs
for a thin cool disk, respectively. The capture rate is essentially a problem of 
loss-cone diffusion-diffusion in angular momentum rather than energy.
By assuming a Salpeter mass function for the stars, \citet{syer} have estimated 
the capture rate in our Galaxy as $\sim 4.8 \times 10^{-5} yr^{-1}$ for main
sequence stars and $\sim 8.5 \times 10^{-6} yr^{-1}$ for red giant stars,
respectively. However, the actual capture rate depends sensitively on the
assumed mass function of stars, the stellar evolution model used, the radius and
mass of the captured star, the black hole mass and the internal dispersion
velocity of stars ($v_s$) around the black hole. For example, \citet{chelu} 
obtained a longer capture time $\sim 10^6$ years, by taking $v_s =10^2 \rm
{km/s}$ and $M_{bh}=2.45 \times 10^6 M_{\odot}$. Therefore the capture time for
a main sequence star with mass $\sim 1M_{\odot}$ could range from several ten thousand
years to several hundred thousand years. The capture time for the more massive
stars is expected to be even longer.  For t$> t_{peak}$, the accretion rate
evolves as, \citet{rees,Phi89}
\begin{eqnarray}
\dot{M}\sim
\frac{1}{3}\frac{M_*}{t_{min}}\left(\frac{t}{t_{min}}\right)^{-5/3}\,\,\,
\label{rees}
\end{eqnarray}
 where  ${M_*}$ and $R_*$  are the mass and the radius of the captured star,
respectively and $t_{peak}\sim 1.59 t_{min}$, where
\begin{equation}
t_{min} \approx 0.2 \left(\frac{M_{\odot}}{M_*}\right)\left(\frac{R_*}{R_{\odot}}\right)^{3/2}\left(\frac{M_{bh}} 
{10^6M_{\odot}}\right)^{1/2}\,\,\mbox{yr}
\end{equation}
is the characteristic time for the debris to return to the pericenter \citep{lu}. The recent
$\it Chandra$ observations of three large amplitude, high-luminosity soft
X-ray flares in AGNs provide strong evidence for the tidal capture events, and
the decrease of X-ray luminosity indeed follows closely 
the above theoretical predictions (e.g. Halpern, Gezari and Komossa 2004).

\citet{sir} analyzed the tidal disruption of stars near a black hole, a process which is a
possible mechanism for the activity of galactic nuclei. They concluded that the
total power released during the destructions of a star is on average $4\cdot
10^{43}$ erg s$^{-1}$. However, the duration of the energy emission due to the
absorption of a star ($\sim 200$ years) is much shorter than the mean time
between two absorption events. Therefore most of the time galactic
nuclei are in a low active state. Furthermore, the maximum energy release must be
much higher than the mean estimate. For example, the highest luminosity
observed in PG 0052+251  \citep[see Table 1 in][]{sir} is $5\times 10^{44}$ erg
s$^{-1}$, which gives a total energy release during 200 years of about $3\times
10^{54}$ erg.

It is very important to know how much of the accretion power will be converted into
the out-flow of relativistic protons. Processes of particle acceleration near
black holes are not well known though several models of particle acceleration
in accretion disks and jets of black holes have been developed \citep[see, e.g.,][]
{kard,heinz,truong,aha05}. For our aims we estimate roughly from
Eq.(\ref{rees}) the energy of disruption which can be transferred to protons.
 In the case of star capture, it is very natural to assume that the resulting jet contains
mainly protons simply because of the dominance of hydrogen in stars. 
We do not know how the jet is formed but if we take the analogy of
Active Galactic Nuclei (AGN), then we can argue that the observed Doppler
factor of AGN is of order 10, and we just assume that the jet coming out from
our galactic black hole is mildly relativistic and its main composition is protons.

\citet{falck} have argued that the conversion efficiency ($\eta_p$) from accretion 
power ($\dot{M}c^2$) into the the energy of jet motion ranges from $10^{-1}$ 
to $10^{-3}$. Integrating the  Eq.(3) and using typical values 
of the parameters, the energy carried away by relativistic protons is estimated as
\begin{eqnarray}
\Delta{E_p} \sim 6 \times 10^{52}(\eta_p/10^{-1}) (M_*/M_{\odot}) \mbox{erg}.
\label{erg}\,\,
\end{eqnarray}

In addition to the accretion power, \citet{chelu} have argued that if the
transient accretion disk can generate a sufficiently strong magnetic field, due to
the instability of the disk, this strong magnetic field can initiate the
Blandford-Znajek process \citep{blaz} to extract rotation energy from the
black hole. They estimate that the maximum energy that can be extracted from a
black hole is given by
\begin{eqnarray}
\Delta{E_{max}} \sim 3 \times 10^{52}A^2 f(A)M_6^2 \mbox{erg}.
\label{bherg}\,\,
\end{eqnarray}
where $A$ is the dimensionless angular momentum of the black hole, $f(A)$ is a
constant for given $A$ and $M_6$ is the black hole mass in units of
$10^6~M_{\odot}$. If the black hole is rotating in maximum angular velocity,
$A=1$, then $f(A)$=1.14.

 The maximum energy in relativistic protons can be estimated from Eq.(5) or
(\ref{bherg}). If a star with the mass about $50~M_\odot$ is captured by
a black hole, it gives an energy in relativistic protons as high as $\sim
10^{54}$ erg.

Thus, we conclude that when eventually a massive star is captured, a huge amount of energy
can be released in the form of relativistic protons during a very short time.
Below we assume that such primary protons interact with the medium gas in the
region, its intensity derived from the gamma-ray data, and produce there secondary gamma-rays
and positrons.

In the summary of these two sections we present the conditions which our
model of annihilation emission by the black hole should satisfy:
\begin{enumerate}
\item The rate of production of thermal positrons at present should be $\sim 10^{43}$
s$^{-1}$;\\
\item If positrons are produced via $p-p$ collisions the total energy of
relativistic protons  must not exceed  $10^{53}-10^{54}$ erg;\\
\item The eruption time of protons is much smaller than any other characteristic time
of the problem. Therefore the production function of protons can be described
as a delta function in time;\\
\item Though positrons are emitted by the black hole - a point-like source-
their spatial distribution when they cool down to thermal energies should extend
over a region with an angular radius $5^o-8^o$ around the Galactic Center;\\
\item If the gamma-ray emission near the Galactic center and the annihilation
emission have a common origin the flux of gamma-rays with energies
$E_\gamma>500$ MeV should be about $F_\gamma\sim 2\cdot 10^{37}$ erg s$^{-1}$.\\
\end{enumerate}

\section{Proton Spectrum}

We take the source function of protons as
\begin{equation}
Q(r,E_p,t)=A(E_p)\delta({\bf r})\delta(t)
\end{equation}

Particles can be accelerated near the hole by a shock wave appearing in the
accreting matter. According to typical shock acceleration models
the momentum spectrum of accelerated particles is a power-law,
with a spectral index $\gamma_0$ between 2-3. If we transform the momentum spectrum to
an energy spectrum, then it has the form
\begin{equation}
\label{s_sp}
 A(E_p)=A_0^\prime{{E_p+M_pc^2}\over{(E_p^2+2M_pc^2E_p)^{(\gamma_0+1)/2}}}
\end{equation}
In fitting the observed diffuse gamma-ray spectrum we should be able to provide some
constraints on $\gamma_0$. However, the error bars of the observed data are
quite large (cf. Fig. 3), and the possible range of $\gamma_0$ is around 2.6-3.1.
In fact the best fit gamma-ray spectral index is 3.1 for photon energies larger
than 2 GeV (cf. Eq.2). However, the exact value of $\gamma_0$ affects
neither the gamma-ray luminosity nor the annihilation flux significantly,
because most of the energy comes from protons with energies of several GeV. Below we take
$\gamma_0 \sim 2.6$ as reference and the constant $A_0^\prime$ will be found
from the gamma-ray data.

The spatial distribution of the protons can easily be derived from the
well-known equation of cosmic ray propagation \citep[see][]{ber90},
\begin{equation}\label{eqp}
{{\partial n_p}\over{\partial t}}-\nabla (D_x \nabla n_p)+{{\partial
}\over{\partial t}}\left({{dE}\over{dt}}n_p\right)+{n_p\over {\tau_p}}=Q({\bf
r},E_p,t)
\end{equation}
Here $dE/dt$ is the term describing particle energy losses and $\tau_p$ is the
characteristic time of catastrophic losses.

In the relativistic energy range the lifetime of protons is determined by $p-p$
collisions, and  equals $\tau_p\simeq 10^{15}$ s for a gas density $n=1$
cm$^{-3}$ . The rate of continuous energy loss is negligible in this energy
range ($dE/dt\simeq0$). The solution of Eq. (\ref{eqp}) in  this case is
similar to the solution of the equation of thermal conductivity

\begin{equation}
n_p(r,E_p,t)=A\exp\left[-t/\tau_p\right]\label{np2}
{{\exp\left[-{{r^2}\over{4D_xt}}\right]}\over{(4\pi D_xt)^{3/2}}}\label{sol_p}
\end{equation}
and describes an almost uniform proton distribution within the sphere of the
radius $r\la \sqrt{D_x t}$.

In the non-relativistic energy range the injection spectrum is transformed by
ionization losses which for a medium density $n=1$ cm$^{-3}$ are significant
at relatively low proton energies. The rate of ionization losses is
\citep{haya, ginz}
\begin{equation}
{{dE}\over{dt}}=-{{2\pi e^4n}\over{mc\beta(E)}}\ln\left({{m^2c^2W_{max}}
\over{4\pi e^2\hbar^{2} n}}\right)\simeq -{a\over\sqrt{E}} \label{iot}
\end{equation}
where $W_{max}$ is the highest energy transmitted to an ambient electron, and
$\beta(E)=v/c$.
 The characteristic time
 of ionization (Coulomb) losses is $\tau_c\sim E/(dE/dt)$. Variations of
 ionization loss times for electrons (dashed-dotted line) and protons (solid line)
 are shown in Fig.\ref{ion}.

The equation for the total number of particles $N(E,t)=4\pi\int\limits_0^\infty
n_p(E,r,t)r^2dr$ can be obtained by integration of Eq.(\ref{eqp}) over the
volume occupied by the particles when $dE/dt\neq 0$ and  is described by
Eq.(\ref{iot}),
\begin{eqnarray}\label{np1}
&&N_p(E_p,t)={2\over 3}{{A_0\sqrt{E_p}}\over{\left({{3at}\over
2}+E_p^{3/2}\right)^{(\gamma_0+2)/3}}}{{\left[\left({{3at}\over
2}+E_p^{3/2}\right)^{2/3}+M_pc^2\right]}\over{\left[\left({{3at}\over
2}+E_p^{3/2}\right)^{2/3}+2M_pc^2\right]^{(\gamma_0+1)/2}}}\exp\left(-\frac{t}{\tau_p}\right)
\end{eqnarray}
 The
total spectrum of protons for an arbitrary constant $A_0$ is shown in
Fig.\ref{np}. We see that at the initial stage the spectrum is evolved under
the influence of ionization loss but after the time $t>\tau_p \sim 10^{15}$ s
the $p-p$ nuclear reactions become also essential.

\section{The Gas Components and Particle Propagation in the Bulge Region}
The gas content in the Galactic center is not well known though one can
find its characteristics in \citet{jean}. For the interstellar medium several
components of the gas  were suggested from radio and X-ray data  \citep[see 
the review of][]{ferr01}. They are: a cold neutral component ($n=10$ cm$^{-3}$,
$T=80$ K), a warm neutral component ($n=0.3$ cm$^{-3}$, $T=8000$ K), a warm ionized
component($n=0.17$ cm$^{-3}$, $T=8000$ K) and a hot component ($n=3\cdot 10^{-3}$
cm$^{-3}$, $T=4.5\cdot 10^5$ K).  The distribution of these components in the
bulge region is strongly nonuniform which complicates matters. The point is that
gamma-rays from $\pi^0$ can be produced by all components of the gas if high energy
protons penetrate there while annihilation photons are produced mainly in the
neutral and ionized warm components where the cross-section for annihilation
increases by several orders of magnitude as compared with the other components.
We do not know whether the propagation of high energy protons is the same  in all
components of the bulge medium. We also do not know the characteristics of secondary
positron propagation: whether they are able to penetrate into dense molecular
clouds or what is their diffusion coefficient.

The Bulge (i.e. the region inside a radius $\sim 230$ pc with a height 45 pc)
contains $7\cdot 10^7~M_\odot$ of hydrogen gas.
 90\% of this mass is trapped in small high density ($\sim 10^4$ cm$^{-3}$) clouds
that gives the average gas density in the bulge central region as high as $\sim
700$ cm$^{-3}$ while the remaining $\sim$10\% is homogeneously distributed with
the average density ~10 cm$^{-3}$. The rest of the gas in the bulge is
contained in a region (like an ellipsoid) with a radius ~1.75 kpc. The H1
mass is equally distribute between cold and warm neutral gas.
The mass of warm and hot ionized gas is about $4\cdot 10^7~M_\odot$. 90\% of
this mass is in the warm phase and 10\% is in the hot phase.

The diffusion coefficient is unknown in the Bulge region. \citet{ber90}
estimated the average value of the diffusion coefficient in the Galactic Disk
from analyses of the cosmic ray flux and obtained a value $\sim 10^{27}$
cm$^2$s$^{-1}$. \citet{jean} presented a very rough estimate of the diffusion
coefficient in the Bulge region using the equation derived by \citet{melr}
\begin{equation}
D\simeq
D_B\left(\frac{r_L}{\lambda_{max}}\right)^{1-\delta}\eta^{-1}\label{mel}
\end{equation}
where $D_B=\frac{1}{3}r_Lv$ is the Bohm diffusion coefficient, $r_L$ and $v$
are the particle gyroradius and velocity, respectively,
$\lambda_{max}\simeq 100$ pc is the maximum scale of turbulence, $\delta=5/3$
for a Kolmogorov turbulence spectrum, and $\eta \sim\delta B^2/B^2\sim 1$ is the
pressure ratio of magnetic fluctuations to the main field.

For a magnetic field strength $B\sim 10^{-5}$G \citep[see, e.g.][]{rosa} the
diffusion coefficient of 30 MeV positron is about $10^{27}$cm$^2$s$^{-1}$.
Though the diffusion coefficient decreases with particle energy, just this value
 determines positron propagation in the Bulge region because the characteristic
 time $\tau_c\propto E^{1}$, i.e. positrons spend most of their lifetime with
 high energy.

The  problem of particle penetration into molecular clouds was analyzed by
\citet{ss,dosh}. In the absence of waves the cosmic ray (CR) flux into clouds should be
\begin{equation}
j(E)\simeq N(E)v
\end{equation}
where $v$ is the particle velocity. However, CRs of a few MeV excite waves
to such a high level that the flux reduces to
\begin{equation}
j(E)\simeq N(E)V_A
\end{equation}
where the Alfvenic velocity $V_A<<v$.

 Therefore, Alfven waves generated outside molecular
clouds by the flux of cosmic rays are effective at excluding cosmic rays below
a few hundred MeV, while particles with higher energies freely penetrate into
the clouds.
This feature plays a very important role in estimating the amount of injection energy of protons for explaining the 
observed gamma-rays and the annihilation emission.

As we see from our analysis several characteristic times determines the
conditions under which the annihilation flux is formed. These times are:
\begin{itemize}
\item The current time from the moment of eruption $t$:\\
\item The time for nuclear $p-p$ reactions (the lifetime of protons) $\tau_p=nv_p\sigma_{pp}$.
For gas densities ranging from 1 to 100 cm$^{-3}$ it ranges from
$10^{15}$s to $10^{13}$s;\\
\item The cooling time of positrons $\tau_c$(the thermalization time)
\begin{equation}
\tau_c\sim \int\limits_{E_{e^+}}^{E_{th}}\frac{dE}{(dE/dt)_c}
\end{equation}
where $(dE/dt)_c$ is the energy cooling rate, which will be discussed in section 7, $E_{e^+}$and $E_{th}$ are the 
injected energy and the thermalization energy respectively. For a density $n$
  we estimate that $\tau_c \sim 3\times 10^{14}(n/1~cm^{-3})^{-1}$ s.\\
\end{itemize}

In the following sections, we investigate models which are characterized by
different ratios between these time scales.

\section{Gamma Rays from $\pi^0$-Decay }

Pions are produced in $p-p$ collisions whose cross-section is well-known
\citep[see, e.g.,][]{steph, aton, derm, dog90, strong}.

The production function of gamma-rays is
\begin{equation}
\label{ng} q_{\pi^{0}} = \eta\int_{E_{p}}N(E_{p})v_pn_{H}d\sigma(E,E_{p})
\end{equation}
where $\eta$ is a factor depending on the chemical composition of cosmic rays
and of the interstellar medium, which for the Galaxy equals 1.6 \citep{steph}, $v_p$
is the velocity of protons and $d\sigma$ is a differential cross-section for
the production of a $\pi^{0}$ of energy E by a cosmic ray proton.

According to \citet{fat} the cross-section is
\begin{equation}
\frac{d\sigma(E_p,E_{\pi^0})}{dE_{\pi^0}} =
\frac{\sigma_0}{E_{\pi^0}}f_{\pi^0}(x)
\end{equation}
where $x=E_{\pi^0}/E_p$, $\sigma_0$ = 32 mbarn and $f_{\pi^0}(x) =
0.67(1-x)^{3.5}+0.5e^{-18x}$. It is important to note that the above formula
is based on the isobar model of Dermer (1986), which is technically appropriate
for a modest range of energies above pion decay threshold. At the highest
energies detected by EGRET, the scaling model is more appropriate and this is
also given in Dermer (1986).

The emissivity of the photons produced by $\pi^{0}$-decay
($q_{\gamma}$) can be calculated from (\citet{steph}):
\begin{equation}\label{eq_gamma_pi_0}
q_\gamma(E_\gamma, t) = 2\int\limits_{E_{\pi min}}^{\infty}
\frac{q_\pi(E_\pi,t)}{\sqrt{E_\pi^2 - m_\pi^2c^4}}dE_\pi
\end{equation}
where $E_\gamma$ and $E_\pi$  are the energy of the emitted photon and the
decaying pion respectively, $E_{\pi min}=E_\gamma+\frac{m_\pi^2c^4}{E_\gamma}$

The spectrum of expected gamma-ray emission is shown in Fig. \ref{gf} together
with the EGRET data. From the observational data we concluded that  at the
present time the constant $A_0$ is
\begin{equation}
A_0=4\times 10^{55} \mbox{GeV$^{\gamma_0-1}$} \label{a0}
\end{equation}
and the total energy of accelerated protons emitting gamma-rays is
\begin{equation}
W_p=\int\limits _{E_p}^\infty E_pN_p(E_p)dE_p\simeq 1.5\times
10^{53}\mbox{erg}\,,
\end{equation}
for a density $n=1$ cm$^{-3}$. If we take the average density of the gas in
the region of gamma-ray emission $n=10$ cm$^{-3}$ we obtain
$W\simeq 10^{52}$ erg for the same production rate of gamma-rays.

 Below we determine whether these protons
are able to generate enough positrons in order to produce the observed flux of
the annihilation line

\section{Production Function of Secondary Electrons}
Inelastic $p-p$ collisions produce two charged pions for every neutral pion.
These charged pions quickly decay into muons, which in turn decay into positrons
and electrons, with a resulting emissivity
\begin{eqnarray}\label{ne}
q_e(E_e)= n_{H}c\frac{m_{\pi}^{2}}{m_{\pi}^{2}-m_{\mu}^{2}}\int\limits _{E_\mu^{min}}^{E_\mu^{max}} dE_\mu 
\frac{dP}{dE_e} \int\limits
_{E_\pi^{min}}^{E_\pi^{max}} \frac{dE_\pi}{\beta_\pi E_\pi}
\int_{E_{th}(E_\pi)} dE_p n_p(E_p) \frac{d\sigma(E_\pi,E_p)}{dE_\pi}
\end{eqnarray}
The integration limits in the above expressions are determined through kinematic
considerations: for example for two-body decay the minimum and maximum energy
in units of the rest mass of the created particles are $\gamma-p$ and $\gamma+p$, where $\gamma$
is the Lorenz-factor of the source particle and $p = \sqrt{\gamma^2-1}$ is its
dimensionless momentum. The lepton distribution from a decaying muon is given by
the three-body decay probability
\begin{eqnarray}
\frac{dP}{dE_e} = \frac{8pc}{\beta_\mu m_\mu^3c^6}\int du \frac{u(u^2\gamma_\mu^2-m_e^2c^4)^{1/2}} {(pc-E_e+u)^2}
\left(3-\frac{4\gamma_\mu u}{m_\mu c^2}\right)
\left(1-\frac{E_e(E_e-u)}{p^2c^2}\right)
\end{eqnarray}
where $u=(E_e-\beta_\mu pc\cos\theta)$, $p$ is the electron momentum and $\gamma$
is the Lorenz factor for the designated particle \citep{fat}.

The production function of $\pi^\pm$ secondary electrons is shown in
Fig.\ref{e_pr}. The secondary $\pi^\pm$ electrons are generated above the
threshold energy $E\ga 30$ MeV. Below this energy secondary electrons are
produced by Coulomb collisions (knock-on electrons). The cross-section of KO
electron production is \citep[see, e.g.,][]{haya, ginz}
\begin{equation}
\sigma_{ko}\simeq 2\pi\left({e^2\over{mc^2}}\right)^2{{mc^4}\over v^2}{1\over
E_e^2}
\end{equation}

The composition of the production flux is an essential function of particle
energy. The electron flux at energies above $\sim 30$ MeV consists of almost
equal parts of positrons and electrons. Below this energy the contribution of
positrons is negligible.

From Eqs. (\ref{np1}), (\ref{a0}) and (\ref{ne}) we can estimate the total
production rate of relativistic positrons at GC at the present moment directly
associated with the current gamma-ray production. It equals
\begin{equation}
 {{dN_{e^+}}\over{dt}}\simeq  10^{41}\mbox{s$^{-1}$}\,.
 \label{pos_pr}
\end{equation}

Below we estimate the number of thermal positrons supplied via $p-p$
collisions after their cooling by Coulomb interactions (ionization losses). The
total electron production function for $\pi^\pm$ and KO electrons is shown in
Fig. \ref{e_pr} where we used the results of \citet{danl}.

The distribution function of these relativistic positrons is described by
Eq.(\ref{eqp}). In the relativistic energy range electrons lose energy by
synchrotron and inverse Compton emission
\begin{eqnarray}
{{d E_e}\over{d t}}=-{{32\pi c}\over
9}\left({e^2\over{mc^2}}\right)^2{c\over{mc^2}}\left(w_{ph}+6.2
~10^{11}{H^2\over{8\pi}}\right)E^2=-\beta_1 E^2
\end{eqnarray}
The characteristic time of energy loss $\tau$ in this case is
\begin{equation}
\xi=\int\limits_{E}^{E_0}{{dE}\over{\beta_1 E^2}}
\end{equation}
Then for the proton spectrum (\ref{sol_p}) and the production function
(\ref{ne}) the spectrum of secondary electrons can be derived from the general
equation (\ref{eqp}), see \citet{syr} and \citet{grat},
\begin{eqnarray}
&&N_e(E_e)={{ 1}\over{ E_e^2}}\int\limits_0^t\theta(t_0)dt_0\label{gen}
\int\limits_0^\infty r_0^2dr_0
{{\exp\left[-{{r_0^2}\over{4D_xt_0}}\right]}\over{(4\pi D_xt_0)^{3/2}}}
\int\limits_0^\eta d\tau{{q_e(\eta-\xi)}\over{(4\pi
D_x\xi)^{3/2}}}\times\\
&&\times\exp\left[-{{({\bf r}-{\bf r}_0)^2}\over{4D_x\tau_0}}\right]\delta
(t-t_0-\tau) \nonumber
\end{eqnarray}
where
\begin{equation}
\eta={1\over{\beta_1 E_e}}\,,
\end{equation}
$q_e$ is derived from Eq. (\ref{ne}), and $\theta(x)$ is the Heaviside
function.

If there were several eruption processes we should sum
over them in order to get a resulting spectrum of positrons.

\section{Evolution of the  Positron Spectrum}
\subsection{Kinetic Equation}
In order to estimate the flux in the $e^+-e^-$ annihilation line we
should calculate how many thermal positrons can be supplied by $p-p$ nuclear
reactions under the influence of energy losses. Similar calculations were
performed by \citet{steck} who analyzed effects of energy losses only.
\citet{cran} calculated the positron density produced in solar flares from the
Fokker-Planck equation when positrons are injected at an energy of 1 MeV.
Recently \citet{guessoum} have studied the lives and deaths of positrons in the
interstellar medium by using  Monte Carlo simulations as well as a binary
reaction rate approach. In our case we should use the generalized Fokker-Planck
equation because positrons are injected at a relativistic energy of about 30
MeV.

The generalized Fokker-Planck equation, describing the particle distribution
function $f$  written in dimensionless variables is \citep[see][]{dog00,liang}

\begin{eqnarray}\label{eq_pos}
\frac{\partial f}{\partial t}+{{vn(\sigma_{an}+\sigma_{ce})}\over\nu_0}
f-q_e(p,{\bf r})= \frac{1}{p^{2}}\frac{\partial} {\partial p}\left[{\cal
A}(p)\frac{\partial f(p)}{\partial p}+{\cal B}(p) f\right]\,,
\end{eqnarray}
where  $q_e$ describes the distribution of sources emitting fast particles.
If we compare this equations with Eq.(\ref{eqp}) for relativistic and
subrelativistic cosmic rays we see that in order to describe evolution of the
positron spectrum from relativistic  to thermal energies we include the terms
in rhs which are responsible for the formation by Coulomb
collisions of the equilibrium Maxwellian
distribution with a given temperature of the background gas.
The injection spectrum is given by the term $q_e$ which is
determined by the production of secondary electrons at energies above 30 MeV.  As
one can see from Fig. \ref{ff1} the change in $f$  at thermal energies
implied by Eq. (\ref{eq_pos})
is then a simple scaling with time, mediated by the
ionization loss timescale. The charge-exchange annihilation processes
describing by the cross-section $\sigma_{ce}$ act for positrons as an absorbing
boundary at low energies, especially for eV temperatures when the cross-section
of annihilation increases by several orders of magnitude \citep{buss,guessoum}.
Therefore, for a constant production of positrons we have in the stationary
case the annihilation rate of thermal positrons exactly equal to $\int
q_e(p)p^2dp$.

In our case this source function describes the production of positrons (shown in
Fig. \ref{e_pr}) with the rate $\sim  10^{41}~e^+$ s$^{-1}$GeV$^{-1}$, the main
part of which is concentrated near energies $E_{e^+}\sim 30-50$ MeV, and
processes of positron annihilation are described by cross-sections
$\sigma_{ce}$ and $\sigma_{an}$ (see section 6.1).

We define non-dimensional variables $p=p/\sqrt{mkT}$ and $t=\hat t\cdot \nu_0$,
 where the characteristic frequency $\nu_0$ is
\begin{equation}
\nu_0={{2\pi \bar{n}_{e}c^{2}r_{e}^{2}m_{e}}\over{\sqrt{m_{e}kT_{x}}}}\,,
\end{equation}
where $\bar{n}_{e}$ is the average electron number density, and $r_e$ is the
classical electron radius $r_{e}=e^{2}/m_{e}c^{2}$, $\bar{n}_{e}= 1$\,cm$^{-3}$
and $T_{x}=2.5$\,eV, $\nu_0\simeq 6.7\times 10^{-12}$ s$^{-1}$.

In Eq.~(\ref{eq_pos}), ${\cal B}$ describes the particle energy gain and loss
and ${\cal A}$ the momentum diffusion due to Coulomb collisions with background
particles,
\begin{equation}
{\cal
A}(p)=p^{2}\left[-\left(\frac{dp}{dt}\right)_{ion}\frac{\gamma}{\sqrt{\gamma^{2}-1}}\sqrt
{\frac{kT_{x}}
{m_{e}c^{2}}}\right]\,,
\end{equation}
\begin{eqnarray}
{\cal B}(p)= p^{2}\left[-\left(\frac{dp}{dt}\right)_{ion}-
\left(\frac{dp}{dt}\right)_{synIC}-\left(\frac{dp}{dt}\right)_{brem}\right]\,,
\end{eqnarray}
 where $\gamma=1+E(p)/m_{e}c^{2}$ is the Lorentz factor.

The $dp/dt$ terms are rates of momentum loss due to ionization, synchrotron
plus inverse Compton (IC) and bremsstrahlung emission respectively (e.g.
Hayakawa 1969). The ionization loss term is
\begin{eqnarray}\label{il}
 \left(\frac{dp}{dt}\right)_{ion} = -\frac{1}{p}\sqrt{p^{2}+
\frac{m_{e}c^{2}}{kT_{x}}}\frac{\gamma}{\sqrt{\gamma^{2}-1}}
 \times \left[\log\left(\frac{E(p)m_{e}c^{2}(\gamma^{2}-1)}
{h^{2}\omega_{p}^{2}\gamma^{2}}\right)+0.43\right]\,,
\end{eqnarray}
where the plasma frequency $\omega_{p}=\sqrt{4\pi e^{2}\bar{n}_{e}/m_{e}}$. The
combined synchrotron and IC loss term is
\begin{eqnarray}
 \left(\frac{dp}{dt}\right)_{synIC} = -{{32}\over 9}\frac{1}{p}\sqrt{p^{2}+
\frac{m_{e}c^{2}}{kT_{x}}} \frac{1}{\nu_{0}}
 \times \frac{\pi
r_{e}^{2}}{\sqrt{m_{e}kT_{x}}}\left[U_{cmb}+U_{mag}\right]\left(\frac{E(p)}
{m_{e}c^{2}}\right)^{2}\,,
\end{eqnarray}
where $U_{cmb}$ is the energy density of background photons, and $U_{mag}$ is
the magnetic energy density. The bremsstrahlung loss term is given by
\begin{eqnarray}
 \left(\frac{dp}{dt}\right)_{brem} = -\frac{1}{p}\sqrt{p^{2}+
\frac{m_{e}c^{2}}{kT_{x}}} \frac{1}{\nu_{0}}\frac{E(p)}{m_{e}c^{2}}
  \frac{\bar{n}_{e}
m_{e}c^{2}e^{2}r_{e}^{2}}{hc\sqrt{m_{e}kT_{x}}}\left[\log\frac{2(E(p)+m_{e}c^{2})}{m_{e}c
^{2}}-\frac{1}{3}\right] \,.
\end{eqnarray}
The corresponding times of  synchrotron+inverse Compton energy losses are shown
in Fig. \ref{ion}. The magnetic field strength is $H=5\times 10^{-6}$G and
the energy density of  background  photons $U_{cmb}=1$ eV cm$^{-3}$.

\subsection{Numerical Calculations of Thermal Positron Numbers for the
Case of Quasi-Stationary Production ($t<\tau_p\sim \tau_c$)} For
calculations of the positron number at different times we used Eq.
(\ref{eq_pos}) describing the evolution of the positron spectrum in both the
relativistic and nonrelativistic energy ranges. The dimensionless time $t=1$
corresponds to $1.5\times 10^{11}$ s. We include into the
equation the process of positron annihilation whose cross-section at different
energies is \citep[see, e.g.,][]{cran}
\begin{eqnarray}\label{san}
&&\sigma_{an}={{\pi r_e^2}\over{\gamma+1}}\left({{\gamma^2+4\gamma
+1}\over{\gamma^2-1}}
\ln\left[\gamma+\sqrt{\gamma^2-1}\right]-{{\gamma+3}\over\sqrt{\gamma^2-1}}\right)
 \end{eqnarray}

If the energy of positrons is below 1 keV and the medium is weakly ionized
(as is the case at temperatures below 10 eV) then we should also include
the charge-exchange
annihilation, whose cross-section is :

\begin{eqnarray}
\sigma_{ce} = 1.2\times 10^{-15}\cdot\left(\frac{E_{e^+}}{E_0}-1\right)
\exp\left(1-\frac{E_{e^+}}{E_0}\right)~\mbox{cm$^2$}
\end{eqnarray}
where $E_{e^+}$ is the positron energy and $E_0$ = 6.8 eV. This equation was
obtained from the experimental data of
 \citet{sperb} with the approximation of \citet{fatuz}. The corresponding
time of positron annihilation at different energies
$\tau_{an}=(nv\sigma_{an})^{-1}$ is shown in Fig. \ref{ion} by a dotted
line.

The production function of positrons is shown in Fig. \ref{e_pr}.  The
injection function can be presented as $q_e(p,t)\propto Q_0(t)\theta({\bar
p}-{\bar p}_0)p^{-\gamma}$ where ${\bar p}_0$ corresponds to the maximum of the
function $q_e(p)$ at the energy $E=30$ MeV. We assume that annihilation
processes take place in the extended medium surrounding the region, where the
injected protons produce relativistic positrons. The time variation of the
injection function is caused by catastrophic and ionization losses of protons.
However, if we consider a time $t<\tau_p$ we can neglect time variations of
the production function corresponding to the almost stationary case of
positron production. We normalized the distribution function
\begin{equation}
\int\limits_0^\infty p^2dp~q_e(p)=Q_{0}
\end{equation}
where $Q_{0}$ can be obtained from Fig. \ref{e_pr} and is about $ 10^{41}$
s$^{-1}$ for $t<\tau_p$.

If we calculate from the kinetic equation (\ref{eq_pos}) the number of thermal
positrons annihilating in the eV temperature medium on the assumption that
they are produced by the same relativistic protons which are responsible for
the observed gamma-ray flux of $2\cdot 10^{37}$ erg s$^{-1}$ then we obtain the
value $2\cdot 10^{41}$e$^+$s$^{-1}$, which is two orders of magnitude less than
necessary to explain the annihilation flux from the Galactic center. Therefore
the stationary model fails to explain the data. Thus, in the framework of
this model we cannot satisfy both conditions 1) and 5) presented at the end
of Section 2.

\section{Nonstationary Model of Positron Production}
\subsection{Positron Spectrum ($t>\tau_p\sim \tau_c $)}
One of the solutions to this problem might be that in the past the positron
production rate was much higher than follows from the current gamma-ray flux
from the GC. Under this condition the production function of positrons calculated
from Eq.(\ref{np1}) is an essential function of time, $q(p,t)$.

In Fig.\ref{ff1} the evolution of the positron
spectrum calculated from Eq. (\ref{eq_pos}) is shown for the parameters of the
interstellar medium:
 the density $n=1$ cm$^{-3}$, the temperature $T=2.5$ eV, the magnetic field strength $H=5 \times 10^{-6}$G
and the energy density of  background  photons $U_{cmb}=1$ eV cm$^{-3}$.

The distribution function of the positrons injected at the initial stage at
 ${\bar p}>{\bar p}_0$ is shifted to the momentum region ${\bar p}<{\bar
p}_0$ under the influence of Coulomb collisions that can be seen in Fig.
\ref{ff1} as a "bunch" of positrons propagating into the region of small
momenta. Later Coulomb collisions start to form the equilibrium Maxwellian
distribution in the range ${\bar E}\leq 10^{-2}$ keV . At a subsequent point of
time Coulomb collisions continue to form the equilibrium distribution and
accumulate positrons in the thermal energy range, as can be seen in Fig.
\ref{ff1} as the increasing Maxwellian distribution. At the final stage when
the source ceases to work the number of thermal positrons decreases
because of annihilation (see Fig. \ref{tnorm}).

We notice that the positron evolution and the number of thermal positrons
depend on the background gas temperature, e.g. in the plasma with the
temperature 100 eV we can produce a much larger amount of positrons because the
annihilation cross-section in this medium is several orders of magnitude
smaller than in a neutral gas (see Fig. \ref{tnorm_long}).

Now we proceed to calculate the intensity of emission due to annihilation of
positrons. Depending on the temperature and density of the ambient medium,
electrons and positrons can annihilate either directly or via formation of positronium
\citep{buss,gues,guessoum}.

The in-flight differential spectrum of the $\gamma$-rays produced by
annihilation of a positron on the ambient electrons with density $n_{e}$
\citep{ahaat,aha}:
\begin{eqnarray}
q_{an}(\varepsilon) = \frac{\pi r_{e}^{2}cn_{e}}{\gamma_{+}p_{+}}\left[
\left(\frac{\varepsilon}{\gamma_{+}+1-\varepsilon} \right. +
\frac{\gamma_{+}+1-\varepsilon}{\varepsilon}\right) + \\
2\left( \frac{1}{\varepsilon}+\frac{1}{\gamma_{+}+1-\varepsilon}\right) -
\left.\left(\frac{1}{\varepsilon}+\frac{1}{\gamma_{+}+1-\varepsilon}\right)^{2} \right] \nonumber
\end{eqnarray}
where $\gamma_{+} = E_{+}/m_{e}c^{2}$ is the Lorentz-factor of the positron, $p_{+} = \sqrt{\gamma_{+}^{2}-1}$ is the 
dimensionless momentum of
positron and $\varepsilon = E/m_{e}c^{2}$ is the dimensionless photon energy. The energy spectrum of in-flight 
annihilation
\begin{equation}
N_{an}(\varepsilon) = \int\limits_\gamma^\infty
q_{an}(\gamma_{0},\varepsilon)n_{e}(\varepsilon)d\gamma_{0}
\end{equation}
The lower integration limit can be obtained from the kinetic equations and
equals
\begin{equation}
\gamma(\varepsilon) = \frac{\varepsilon^{2} + (\varepsilon-1)^{2}}{2\varepsilon-1}
\end{equation}

The energy spectrum of annihilation photons produced by the charge-exchange
process is \citep{buss}
\begin{eqnarray}
N_{ce}(\varepsilon) = \int\limits_E^\infty
dE_{0}n_{H}n_{e}(E_{0})\sigma_{ce}(E_{0})v \div [(E_{0}-6.8 eV)(E_{0}-6.8 eV +
4\varepsilon)]^{1/2}
\end{eqnarray}
where $E = (\varepsilon-\varepsilon_{0})^{2}/\varepsilon + 6.8 eV$,
$\varepsilon_{0}$ - the rest mass of positronium, $n_{H}$ - the density of
neutral hydrogen.

To fit the INTEGRAL data \citep{chur_INTEGRAL}, shown at Fig. \ref{tflux}, we
used the following medium parameters: plasma temperature 2.5 eV, plasma density
1 $cm^{-3}$, degree of ionization 5\%  \citep[see, e.g.,][]{pik}. The plasma
temperature affects the width of the annihilation line. In order to fit the
INTEGRAL data for energies at and below 511 keV we need a temperature in the
region 2-5 eV.

The line emission at these temperatures is due to the two photon decay of the
charge-exchange  process. The emission below 511 keV is generated by three
photon decay of positronium. The process of in-flight annihilation in a low
temperature medium is negligible because  the time for thermalization in this
medium is smaller than the characteristic time of the in-flight annihilation.
This process is essential in the hot component of the interstellar medium, about
20\% of which is filled by a hot plasma  with the parameters $T=100$ eV and
$n=3\times 10^{-3}$ cm$^{-3}$. However, as one can see from Fig. \ref{an} the
flux of the annihilation line produced in the hot interstellar plasma by
in-flight annihilation is negligible  in comparison with emission in a cold
plasma.

The temporal variations of the gamma-ray and the annihilation fluxes are shown
in Fig. \ref{2em}. The maximum of the fluxes are taken as unity. We see that
the gamma-ray flux is always decreasing with time, while the annihilation flux
reaches its maximum a long time after the proton eruption ($\sim 3\times 10^{14}$
s) and then gradually decreases. We can estimate the necessary input proton
energy without going into the technical calculation details. The current
observed energy flux of gamma-rays is $2\times 10^{37}$erg s$^{-1}$ and the
current annihilation rate of thermal positrons is $10^{43}$ positron s$^{-1}$,
corresponds to a higher proton collision rate in the past.  From Fig.
\ref{2em}, one can see that this situation is realized if  we are observing
these fluxes at a time
 $t>\tau_p$ when the current gamma-ray intensity is at least
three order of magnitude less than its maximum, while the current annihilation
flux falls from its maximum value by less than one order of magnitude (just
because of the delay effect)}. In this case, the necessary input proton energy
is at least $\sim 10^{55}$erg. Of course, this estimate sensitively depends on
the number density. Even if we take the average density of the gas in the bulge
as high as $n=100$ cm$^{-3}$ (see Section 4) we get a value of the energy
necessary to produce the gamma-ray and annihilation fluxes about $W_p\sim
10^{53}$ erg. This amount of energy may be able to be supplied by a
very massive star accreting onto the black hole (see Eq. \ref{erg}) or
by extracting the rotation energy of the black hole (see Eq. 6). However, we may not
be able to obtain a distribution of the annihilation line extending to
several hundred pc from the GC because  the characteristic length of diffusion
propagation is about 100 pc for this high gas density. Thus, in the
framework of this model we have problems with the conditions 2) or 4) from
Section 2. Nevertheless, we conclude that this model is marginally acceptable if
a $ 50~ M_\odot$ massive star was captured by the black hole about hundred
million years ago.

\section{Spatially Non-Uniform Model ($\tau_p<t\sim\tau_c$)}
To solve the possible problem of energy excess we take into account that
the background gas is heterogeneously distributed  in the bulge (see Section
4). According to \citet{jean} hydrogen gas in the nuclear bulge  is trapped in
small high density molecular clouds. Penetration of cosmic rays into molecular
clouds was analyzed in a number of papers \citep[see, e.g.,][]{ces,dosh,pado}.
From the EGRET observations it is known that GeV cosmic ray protons freely
penetrate into molecular clouds \citep{dig} while it is almost unknown how MeV
electrons interact with the clouds. However, \citet{ss} showed that the flux of
cosmic rays below a few hundred MeV  might be completely
excluded from molecular clouds.

In our case it means that relativistic protons propagate in the medium with the
average gas density about n=30 - 1000 cm$^{-3}$ (depending on their propagation
distance from the bulge center) while the secondary positrons ejected from the
clouds propagate in the intercloud medium only, where the average gas density is
about $n\simeq 1-10 ~cm^{-3}$.

For protons it means that they fill the region of a radius $<100$ pc (if the
diffusion coefficient equals the average in the Disk, $D\sim 10^{27}$
cm$^2$s$^{-1}$) during their lifetime $\tau_p \sim (1-5)\times 10^{13}$ s while
electrons can fill a sphere of radius $\sim 400$ pc over the period of time
of their Coulomb cooling ($\sim 3\times 10^{14}$ s for $n=1$ cm$^{-3}$).

The time variations of the gamma-ray and annihilation emission are shown in
Fig. \ref{var_30} for an average density for protons  $n=30$ cm$^{-3}$.
As is to be expected the gamma-ray flux drops rapidly from its initial value
while the annihilation flux reaches its maximum $\sim 3\times 10^{14}$ s after
the eruption. The energy of relativistic protons  necessary to get the peak
production of thermal positrons of  about $10^{43}$ e$^+$s$^{-1}$ is about $7\cdot
10^{53}$ erg. The energy in relativistic protons changes weakly if we increase
further the gas density. Thus, for a density $n=100$ cm$^{-3}$ the necessary
energy of the protons is $5\cdot 10^{53}$ erg. These injected proton energies
are compatible with the estimates in Eq.(5) or Eq.(6) for a massive star capture.

Besides, if the average density of the gas in the medium traversed by
positrons is a little bit higher than 1 cm$^{-3}$ (e.g, 3 cm$^{-3}$) than we
obtain slightly smaller values for the thermolization time ($\sim 10^{14}$s) as
well as the necessary energy output of protons ($\sim 10^{53}$ erg) that makes
our estimates even more acceptable.

 Then, we can assume that the gamma-ray and the annihilation
fluxes from the Galactic center belong to different accretion events.
Gamma-rays were generated relatively recently ($\sim 10^{13}$ s ago) when  a
one solar mass star was captured by the black hole. Indeed the necessary amount
of energy in injected protons is about $\ga 10^{51}-10^{52}$ erg (see Section
5). In this respect it is clear why this emission was observed by EGRET as 
point-like. The average length of proton propagation is  $\la 100$ pc for the
diffusion coefficient $D=10^{27}$ cm$^2$s$^{-1}$. If the annihilation line
emission was produced $\sim 10^{14}$ s ago by a previous event of a 30
M$_\odot$ star capture, with an energy release of $\ga 10^{53}$ erg in protons,
then it  appears as an extended source long after the time when the gamma-ray
emission produced by this capture died out. Thus, this model satisfies the
conditions formulated in Section 2 though the  annihilation and gamma-ray
fluxes observed from the GC are not connected with each other in this case.

\section{Continuous stellar captures}
In the framework of the non-uniform model we can estimate the minimum
energy at relativistic protons  in order to produce in the current time the
thermal positron rate of about $10^{43}$e$^+$s$^{-1}$ and not contradict the
gamma-ray observations. We remember that the present gamma-ray flux from GC is
about $2\cdot 10^{37}$ erg $^{-1}$ and it is seen as a point-like object.
We remember also that the gas distribution is unknown but that the main part of the
gas is concentrated inside the central region of the bulge where the its
density is as high as $\sim 1000$ cm$^{-3}$  \citep[see][]{jean}.  If
relativistic protons freely penetrate into the clouds the
average density  $n\sim 1000$ cm$^{-3}$.  We define this density as $n_{pp}$.

The diffuse gas is distributed in the intercloud medium with an average
density of 10 cm$^{-3}$ and it may reach at the central regions the value of $30
-60$ cm$^{-3}$. If positrons do not penetrate into the cloud then the gas
density of the medium in which positrons are cooled down by the Coulomb
collisions is determined by these values. Below, we define this gas density as
$n_{cc}$. Let us estimate the minimum energy of primary protons necessary to
produce the annihilation flux as a function of $n_{pp}$ and $n_{cc}$.

In the framework of the nonuniform model the initial energy of protons
necessary to produce the annihilation emission  depending on $n_{pp}$ and
$n_{cc}$ is varying in the following way:
\begin{table}[h]\centering
\caption{Injected Proton Energy} \label{t1}
\begin{tabular}{c c c c}\\
\hline\hline
$n_{pp}$(cm$^{-3}$)&$n_{cc}$(cm$^{-3}$)&$W$ (erg)\\
1&1&$10^{55}$\\
30&1&$7\cdot 10^{53}$\\
100&1&$5\cdot 10^{53}$\\
100&10&$2\cdot 10^{53}$\\
1000&1&$2\cdot 10^{53}$\\
1000&10&$1.5\cdot 10^{53}$\\
1000&30&$4.5\cdot 10^{52}$\\
1000&60&$3\cdot 10^{52}$\\
\hline
\end{tabular}
\end{table}
It is clearly seen that when we increase the value of $n_{pp}$ the output energy
of protons decreases because the production rate of relativistic positrons
is proportional to the gas density
\begin{equation}
F_{e^+}=\int N_p(E_p)\sigma_{pp}n_{pp}dE_p \label{pos}
\end{equation}
However as follows from Table 1 the proportionality between $W_p$ and $n_{pp}$
is not linear as it would seem from Eq.(\ref{pos}). From Eq.(\ref{pos})  one
can see that the higher the density $n_{pp}$ the shorter is the injection impulse
of positrons which is determined by the lifetime of protons
$\tau_p=n_{pp}\sigma_{pp}c$.  The necessary energy output of protons as a
function of the ratio $n_{pp}/n_{cc}$ is shown in Fig. \ref{npp}. In the
figure we take the density $n_{cc}=1$ cm$^{-3}$ and $W_0$ is the proton energy
when $n_{pp}/n_{cc}=1$. The variation of the proton injected energy $W_p$ can
easily be understood from the solution of the diffusion equation. When
$n_{pp}/n_{cc}=1$ we have $\tau_p\ga\tau_c$ and the energy $W_p$ at this
density for a fixed flux of thermal positrons at the level $F_{e^+}\sim
10^{43}$e$^+$s$^{-1}$ is of the order of $W_p\propto \tau_p\cdot F{_e^+}$ i.e
it decreases as $W\propto 1/n_{pp}$. However when $n_{pp}\gg n_{cc}$ the
impulse of proton injection is very short $\tau_p\ll\tau_c$ (almost
delta-function injection) and we  obtain $W_p\propto \tau_c\cdot F{_e^+}$, i.e
$W_p$ is independent of $n_{pp}$ as one clearly sees in Fig. \ref{npp}. The time
$\tau_c$ cannot be too short otherwise the gamma-ray flux would not drop down
below the observed gamma-ray flux from the GC (see Fig. \ref{var_30}). Variations
of the gamma-ray flux for a fixed annihilation flux of thermal positrons of
$10^{43}$ ph s$^{-1}$  as a function of the ratio $n_{pp}/n_{cc}$ is shown in
Fig. \ref{g_npp}. From this figure we see that this ratio cannot be below
$n_{pp}/n_{cc}=17$ otherwise the peaks of gamma-rays and the annihilation
emission are so close to each other that the flux of gamma-rays does not have
enough time to drop below the observed limit.

From this restriction we conclude that the injected proton energy cannot be
less than $\sim 3\times 10^{52}$ erg even for $n_{pp}=1000$ cm$^{-3}$ while
$n_{cc}=60$ cm$^{-3}$. This gives the average distance of proton propagation
by diffusion of about 20 pc and about 100 pc for positrons during their
lifetime, which is less than necessary, but the diffusion coefficient may be
spatially dependent such that it gives a more extended positron spatial distribution.
Under this assumption even the capture of 1 $M_\odot$ star can give the necessary
energy in protons as follows from Eq.(5) or Eq.(\ref{bherg}), and the process of
annihilation emission is not an exceptional event as we had in the previous
models when a massive star capture was needed to supply the necessary energy in
relativistic protons.

Since there are so many stars at the GC, there is very good
reason to believe that the capture of stars can continuously take place from
time to time. If there are several successive eruptions near the black hole then the
annihilation flux is fluctuating near a certain level of emission as is
shown in Fig. \ref{ncu3}. In the calculations we take $n_{cc} = 60$ cm$^{-3}$
and $n_{pp} = 1000$ cm$^{-3}$ which give $\tau_c \sim 5\times 10^{12}$ s and
$\tau_p\sim 7\times 10^{11}$ s respectively. The time between two successive
eruption equals $\sim 8\times 10^{12}$ s. When the variations of the energy fluxes
of gamma-rays and the photon flux of the annihilation emission are normalized
relative to their maximum their values, which for gamma-rays is $3\times
10^{39}$ erg s$^{-1}$ and for the annihilation flux is $10^{43}$ ph s$^{-1}$,
the average injected proton energy is $\sim 8\times 10^{52}$ erg.

\section{TeV Gamma-Rays}

It is very interesting to find other confirmations of eruption processes from
observed data. In this respect, the new data in the TeV gamma-ray range are of
strong interest.

Recently, TeV $\gamma$-ray emission at a luminosity higher than $10^{35}$ ergs
s$^{-1}$ from the direction of the galactic center has been reported by three
independent groups, Whipple \citep{kosa}, CANGAROO \citep{tsu}, and HESS
\citep{aha04}. The most plausible candidates suggested for this emission
include the black hole Sgr $A^*$ \citep{aha05} and the compact and powerful
young supernova remnant (SNR) Sgr A East \citep{crock}. The angular scale of
the TeV source was determined by HESS to be about a few arc-minutes, indicating
that this $\gamma$-ray source is located in the central $\sim 10\,pc$ region
\citep{aha04}. The gamma-ray spectrum observed by HESS can be described by $F(
E_{\gamma})=(2.5\pm0.21)\times 10^{-12}E_{\gamma}^{-
2.21}\,\,ph/(cm^2\,s\,TeV)$, which is consistent with decay of $\pi^{\circ}$
produced by p-p collisions with the initial spectrum of the injection protons
as $Q(E)=Q_0E^{-2.2}\exp({-E/E_0})$, where the cutoff energy is
$E_0=10^{15}\,eV$ \citep{aha05}. In comparing with the diffuse gamma-ray spectrum
detected by EGRET (cf. Eq. \ref{fgamma}), the TeV spectrum is flatter. If both
GeV and TeV gamma-rays result from the decay of $\pi^{\circ}$ which are produced by p-p
collisions, it appears that these two different energy gamma-ray spectra must
be produced by two different groups of protons. Furthermore, the protons
responsible for the diffuse GeV gamma-ray emission must be ejected by Sgr A$^*$
much earlier than those responsible for the TeV gamma-rays because their angular sizes are
so much different. In fact, these two requirements are consistent with our
model considered here, i.e. the galactic black hole is activated by capturing
stars.

If we assume that positrons are produced via $p-p$ collisions and these protons 
are emitted from the galactic black hole, by taking the typical diffusion coefficient (D$\sim
10^{27}cm^2 s^{-1}$), we can estimate that these relativistic protons should be
emitted at a earlier time $t_{diff}\sim \frac{(0.5kpc)^2}{6\times 10^{27}cm^2/s}\sim 3\times 10^{14}$s,
which is consistent with the requirement of the annihilation line. Again if we
assume that TeV photons also result from the decay of neutral pions produced by
p-p collisions and that these protons are emitted from the galactic black hole, then
taking the same diffusion coefficient, we obtain that these protons are emitted
at a earlier time $t_{diff}\sim \frac{(10pc)^2}{6\times 10^{27}cm^2/s}\sim 5\times 10^3$years, which is
shorter than the typical capture time of a main sequence star by a factor of
60.
However, D$\sim 10^{27}cm^2 s^{-1}$ is the mean diffusion
coefficient for the Galaxy obtained by cosmic ray measurements \citep{ber90}.
The diffusion coefficient near
the Galactic center can be significantly smaller. For example, the Bohm
diffusion time can be estimated as $t_{Bohm}\sim
(\frac{\lambda}{r_L})^2\frac{r_L}{6c}$ where $\lambda$ is the diffusion distance,
$r_L\approx E_p/eB$ is the Larmor radius, $E_p$ is the proton energy for
producing TeV photons and $B$ is the local magnetic field.  Using $\lambda \sim
$10 pc, $E_p\sim$3TeV and $B\sim 10^{-5}$G, we obtain $t_{Bohm}\sim
2\times 10^5$years, which is similar to the requirement of the continuous capture model.
Taking the observed TeV luminosity $\sim 10^{35}$erg/s that
implies the injected proton energy must be at least $\sim 10^{51}$erg
(Aharonian et al. 2004), which is consistent with our estimation of energy
release for capturing a 1M$_{\odot}$ star (cf. Eq. 5). Furthermore, the
injected proton index for TeV gamma-rays is $\sim 2.2$ but the injected proton
index for the diffuse GeV gamma-rays is $\sim 2.6$ in our model calculation. It
appears that they have different origins. However, this discrepancy can be
explained by the following possibilities. (1) The simplest explanation may be
that the injected proton spectrum can vary from injection to injection. (2) The
most realistic explanation should be due to propagation effects and leakage
effects which occur in the Galaxy where the injection spectrum with the
spectral index as $Q(E)\propto E^{-(2.0-2.2)}$ steepens due to energy
variations of the spatial diffusion coefficient, as $D\propto E^{-0.6}$
\citep[see][]{ber90}.

\citet{lu} have suggested that the TeV emission from the GC resulting from a
powerful relativistic proton jet emitted by Sgr A$^*$ about $\sim 10^4$years ago,
which result from the capture of a red giant star. However, their conclusion
is based on a comparison of the diffusion length of the relativistic protons
with the upper limit on the size of the TeV source. It should be noticed that the
lack of information beyond 10 pc does not mean an absence of such an extended
component. Detection of an extended region with a weaker signal is very
difficult. In this model we predict that when more TeV data are available
the TeV spectrum may show a spatial dependence. In particular, the TeV signals
from a more extended region should exhibit a steeper spectrum if the propagation
and leakage effects are indeed the origins of the spectral discrepancy between GeV
gamma-rays and TeV gamma-rays.

\section{Discussion}
We suggest that the capture of stars by the Galactic black hole is a natural energy mechanism
for providing relativistic protons with a typical energy of $\sim 10^{52-53}$erg.
An essential point of our analysis is the non-stationary processes and the non-uniform density
conditions of emission regions.

The duration of energy emission as well as of the existence of relativistic
protons  after accretion is much shorter than the mean
interval between two accretion events.  Once ejected by the central black hole,
relativistic protons propagate by diffusion into the interstellar medium
surrounding the hole occupying a more and more extended region around the
Galactic Center. Collisions of these protons with the ambient gas result in
secondary gamma-rays and relativistic positrons with energies $E>30$ MeV. The
total gamma-ray flux  is almost constant during the lifetime of the protons but
then it decays exponentionally with the characteristic time dependent on the
gas density.
 Secondary relativistic
positrons produced immediately via $p-p$ collisions are almost unseen at the first
stage after the eruption because the annihilation flux of fast positrons is
almost negligible. These positrons are cooled down by Coulomb collisions with
the ambient gas with the characteristic time for Coulomb collisions. Only when
positrons reach an energy of about several eV they effectively annihilate in the
warm neutral and ionized phases of the interstellar medium where the
cross-section for annihilation increases by several orders of magnitude
\citep[see][]{buss,gues,guessoum} and results in an  increase of the
annihilation emission. The maximum of the annihilation emission is reached long
after the eruption moment. At that time positrons fill an extended region
around the Galactic center by diffusion and  therefore in all cases the
annihilation emission can be seen as an extended region around the Galactic
center, the annihilation flux is never seen as  a point-like central source. On
the other hand, the high energy gamma-rays are produced from the very beginning
of the eruption and can be seen in the first stages after the eruption as a
point-like source.

Our calculations  show that it is marginally possible that the gamma-rays
coming from the GC and the annihilation emission belong to the same eruption
process provided the injected proton energy is $\sim 10^{54}$ erg, this might be
possible theoretically if we take model parameters to their extreme values.  An
alternative assumption that the currently observed gamma-rays from the GC and the
currently observed annihilation emission belong to at least two different
injections seems to be more attractive. The first one belong to a recent
capture of a one solar mass star with the injected energy of about $3\times
10^{51}$ erg while the second one is a consequence of a previous capture of a
much more massive star with an energy release as high as $10^{53}$ erg. This
does not contradict estimations from theoretical models if we assume 
capture of a massive star.

Furthermore,  if processes of secondary particle production occur very
close to the GC where the gas density is about 1000 cm$^{-3}$ then even a 1
$M_\odot$ star capture can provide the necessary energy of protons, which is in
this case $\sim 3\cdot 10^{52}$ erg. The GC is a region with a high concentration of
stars, the capture of a star should continue to take place from time to time.
Although the characteristic capture time scale is not known, TeV emission from
the GC suggests that it should be $>10^5$yr. If this is true, assuming the
positrons cannot penetrate the cloud, so that the cooling time is longer than the
$p-p$ collision time, our calculations indicate that the positron annihilation
rate is more or less constant whereas the current gamma-ray flux is measured
somewhere between two proton injection events. This naturally explains why
the annihilation rate is about two orders of magnitude higher than follows
from the current emission rate of gamma-ray photons.

\begin{acknowledgements}
The authors are grateful A.Aharonyan, E.Churazov, Y.F. Huang , Y. Lu and the anonymous referee
for very useful discussions and comments. We also thank P.K. MacKeown for this critical reading of the manuscript. 
VAD is also grateful to T. Harko and Anisia Tang for their
consultations in carrying out the numerical calculations. This work is
supported by a RGC grant of Hong Kong Government and by the grant of a
President of the Russian Federation "Scientific School of Academician
V.L.Ginzburg".
\end{acknowledgements}

\bibliographystyle{aa}
\begin{figure}[ht]
\plotone{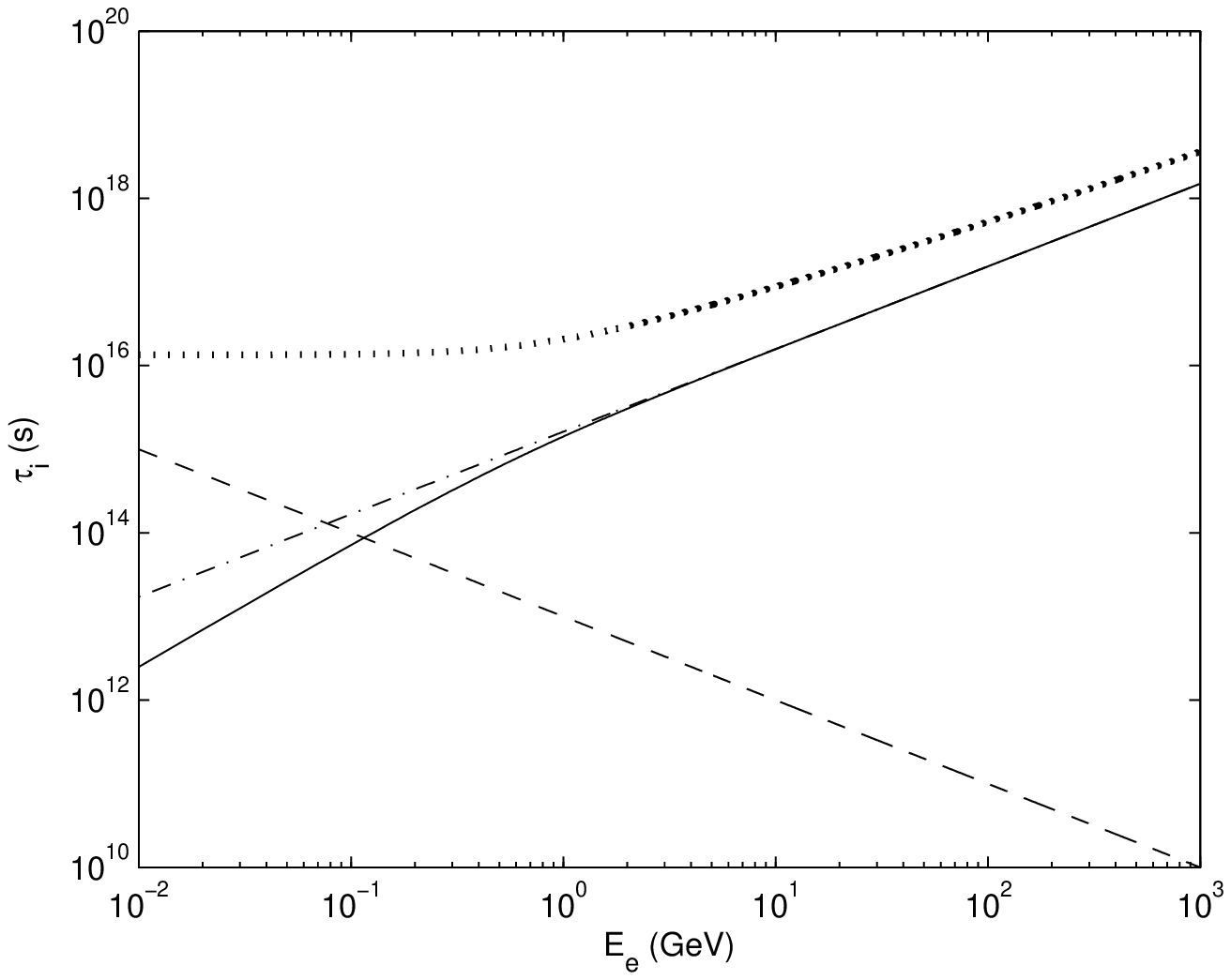} \caption{Variations of
 life-times due to ionization losses when $n=1$ cm$^{-3}$: for electrons (dashed-dotted line) and
 protons (solid line), due to synchrotron ($H=5\times 10^{-6}$ G) and inverse Compton ($w_{ph}=1$ eV cm$^{-3}$)
 energy losses for relativistic electrons (dashed line), and due to direct annihilation (dotted line).
 } \label{ion}
\end{figure}
\newpage

\begin{figure}[ht]
\plotone{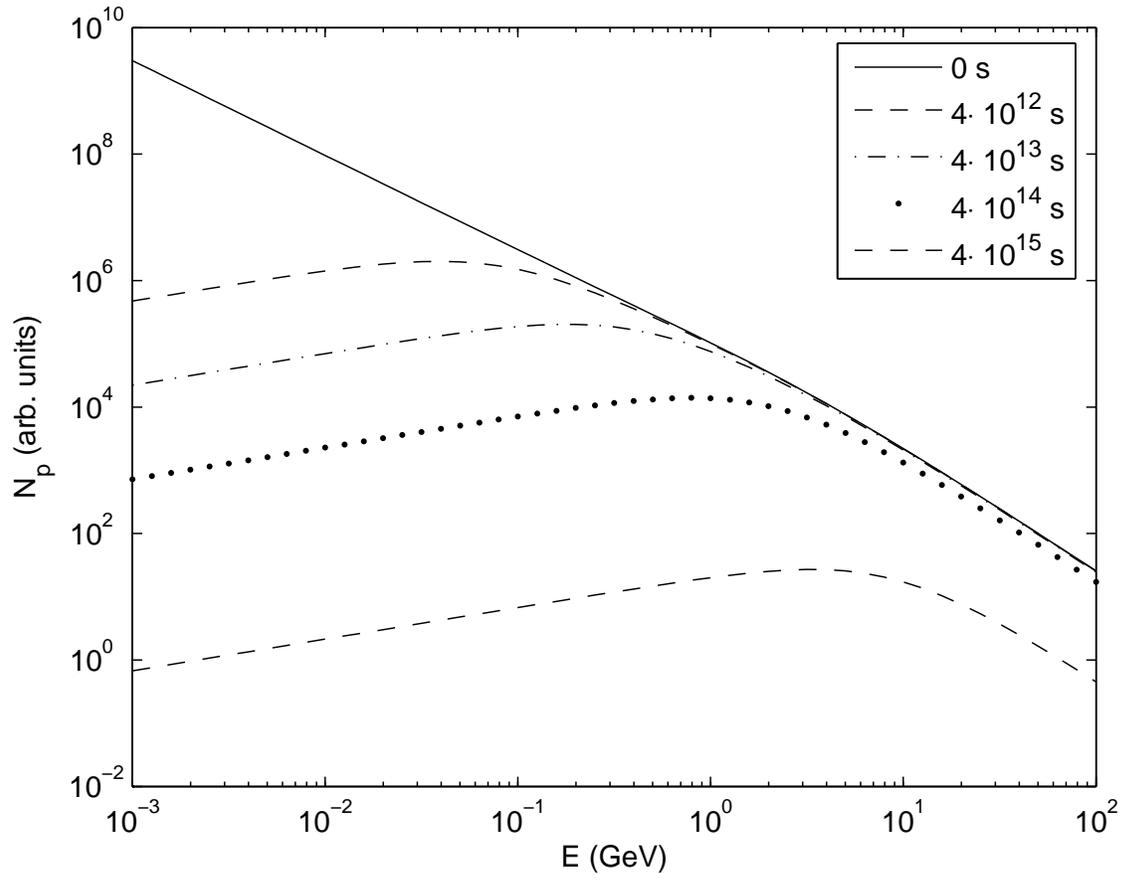} \caption{The spectrum of protons necessary at different
moments of time for an ambient plasma density $n=1$ cm$^{-3}$.} \label{np}
\end{figure}

\newpage

\begin{figure}[ht]
\plotone{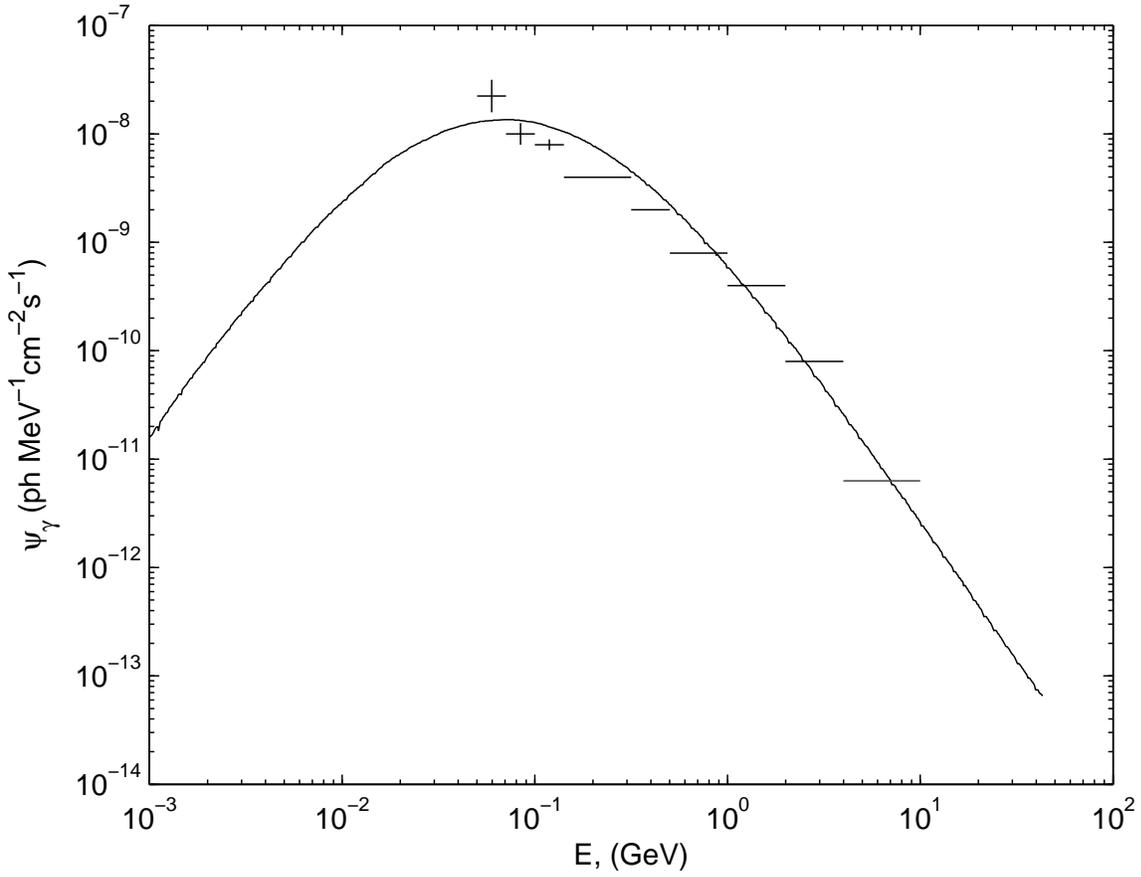} \caption{The flux of gamma-ray emission in the direction of
the Galactic center measured with EGRET and the results of numerical
calculations.} \label{gf}
\end{figure}
\newpage

\begin{figure}[ht]
\plotone{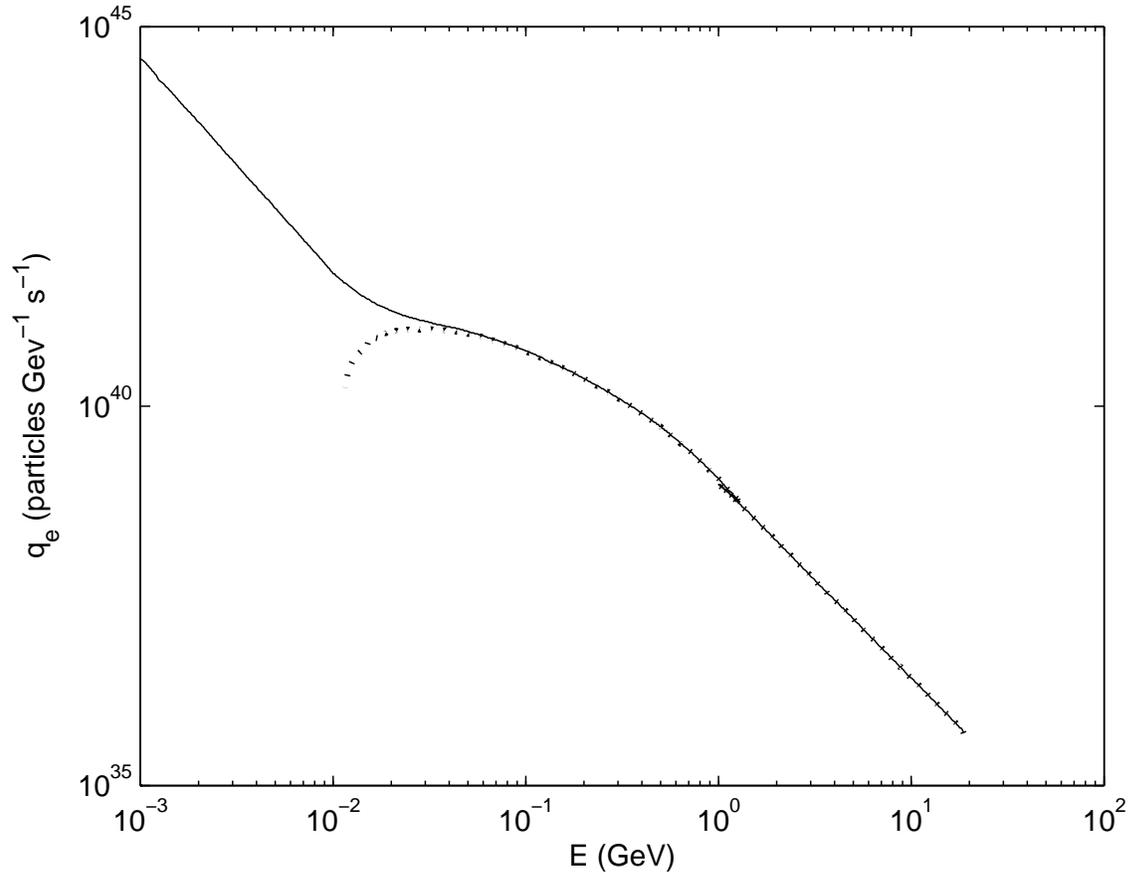}
\caption{The total production function of secondary electrons from $\pi^\pm$-decay and Coulomb collisions (solid 
line) and the production function
of positrons (dotted line).} \label{e_pr}
\end{figure}
\newpage

\begin{figure}[ht]
\plotone{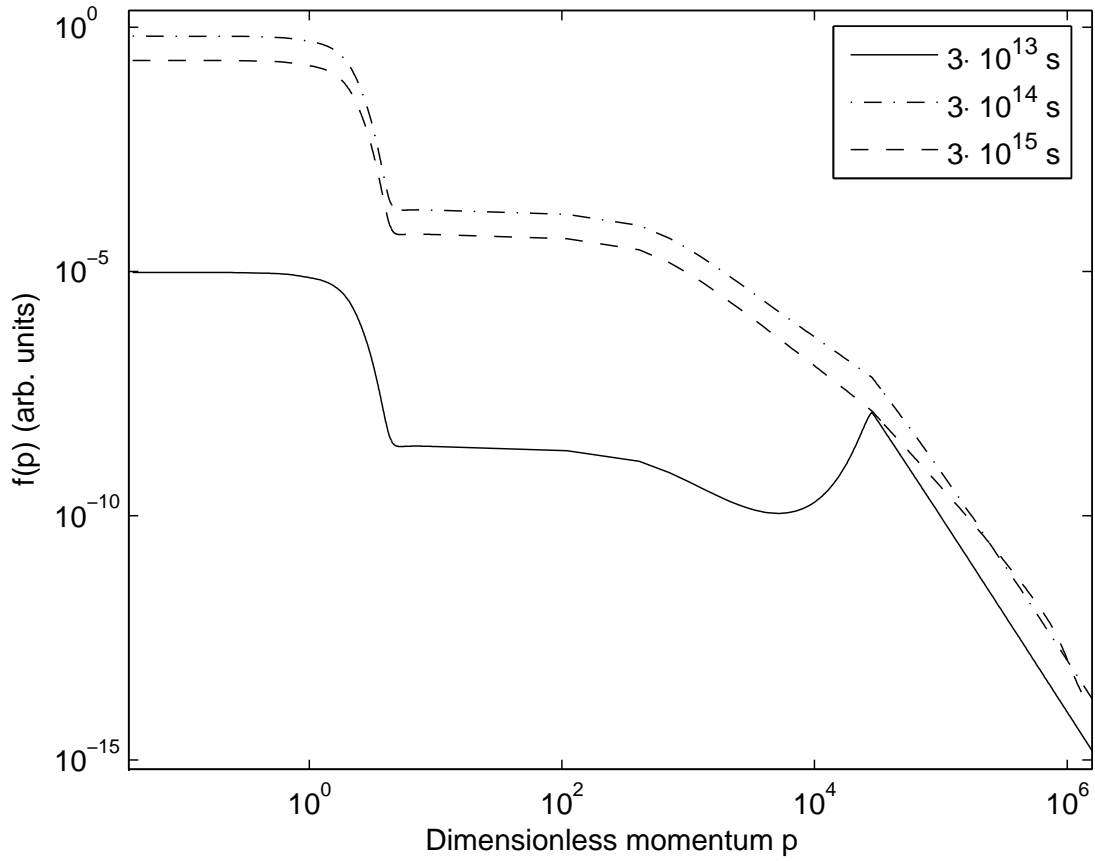} \caption{Time variation of the positron
distribution function $f(p)$ for a gas temperature $T=2.5$ eV. The dimensionless
momentum is defined as $p/\sqrt{m_ekT}$ so the dimensionless
momentum at unity corresponds to a positron with energy $\sim k$T. } \label{ff1}
\end{figure}

\newpage
\begin{figure}[ht]
\plotone{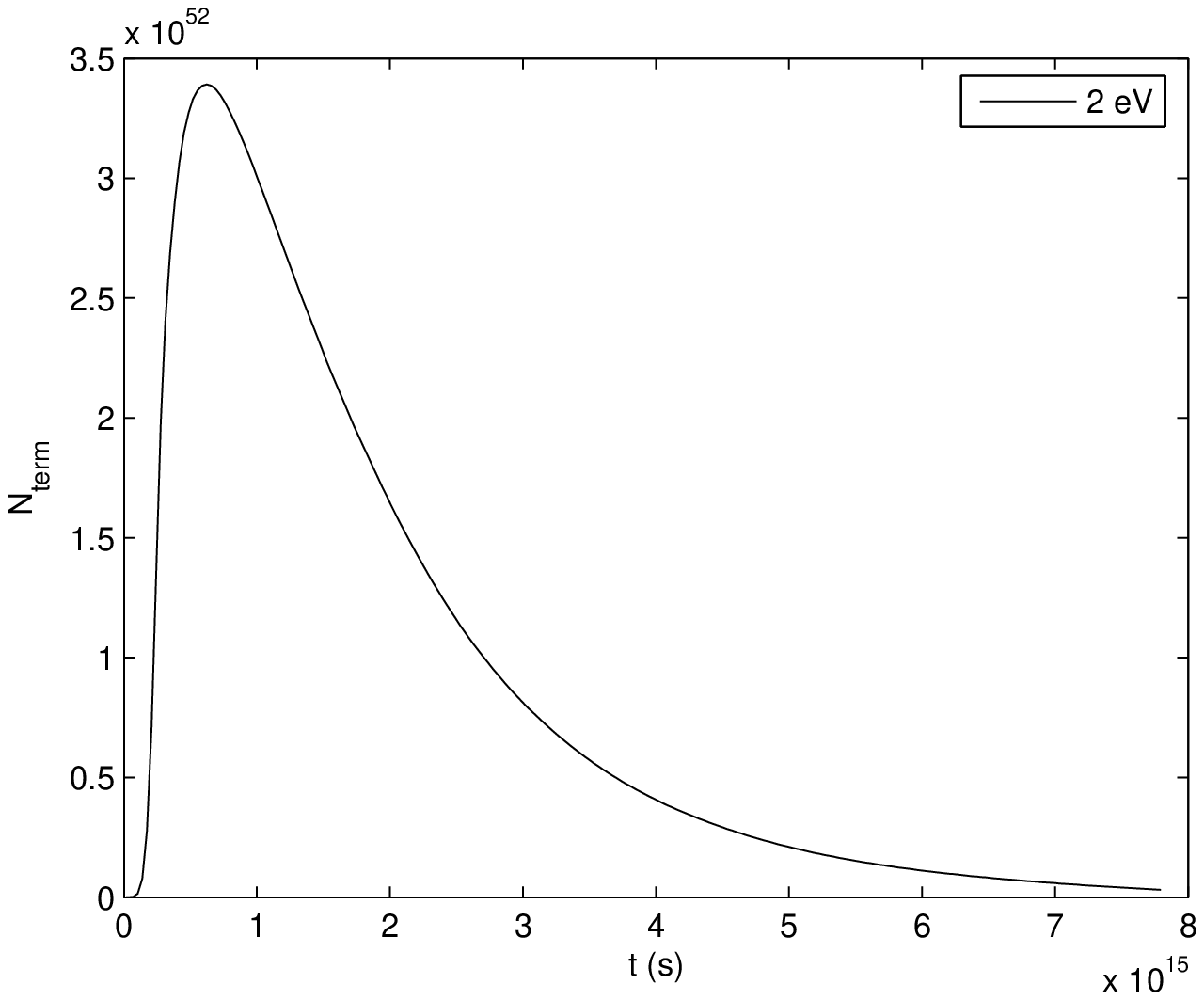} \caption{Time variation of the amount of thermal positrons
produced by relativistic protons injected $4\times 10^{15}$ s ago in the gas
with a temperature $T=2.5$ eV and  a density $n=1$ cm$^{-3}$. }
\label{tnorm}
\end{figure}

\newpage
\begin{figure}[ht]
\plotone{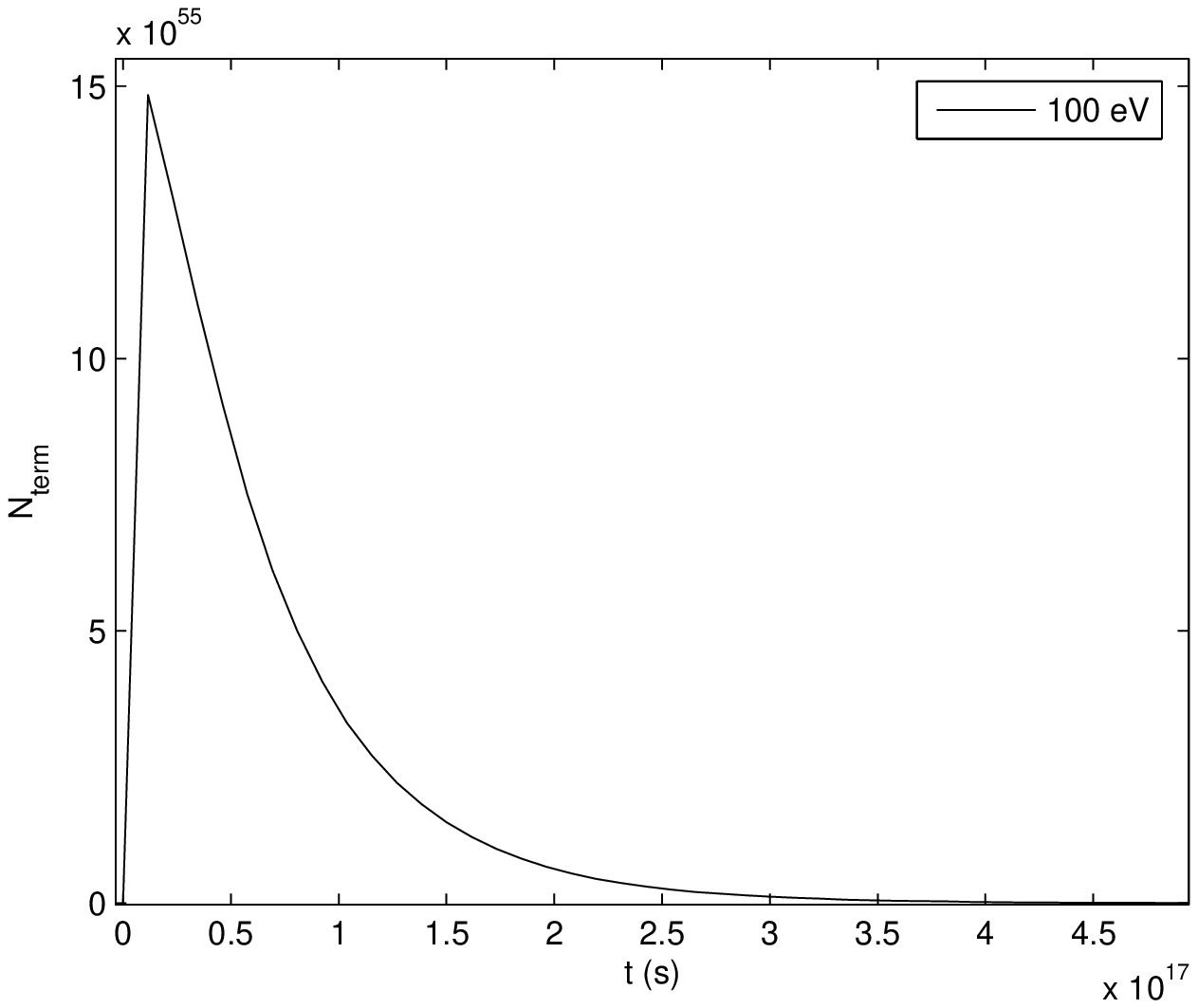} \caption{Time variation of the amount of thermal
positrons produced by relativistic protons injected $4\times 10^{15}$ s ago in
the gas with a temperature $T=100$ eV and  a density $n=3\times 10^{-3}$
cm$^{-3}$.} \label{tnorm_long}
\end{figure}

\newpage
\begin{figure}[ht]
\plotone{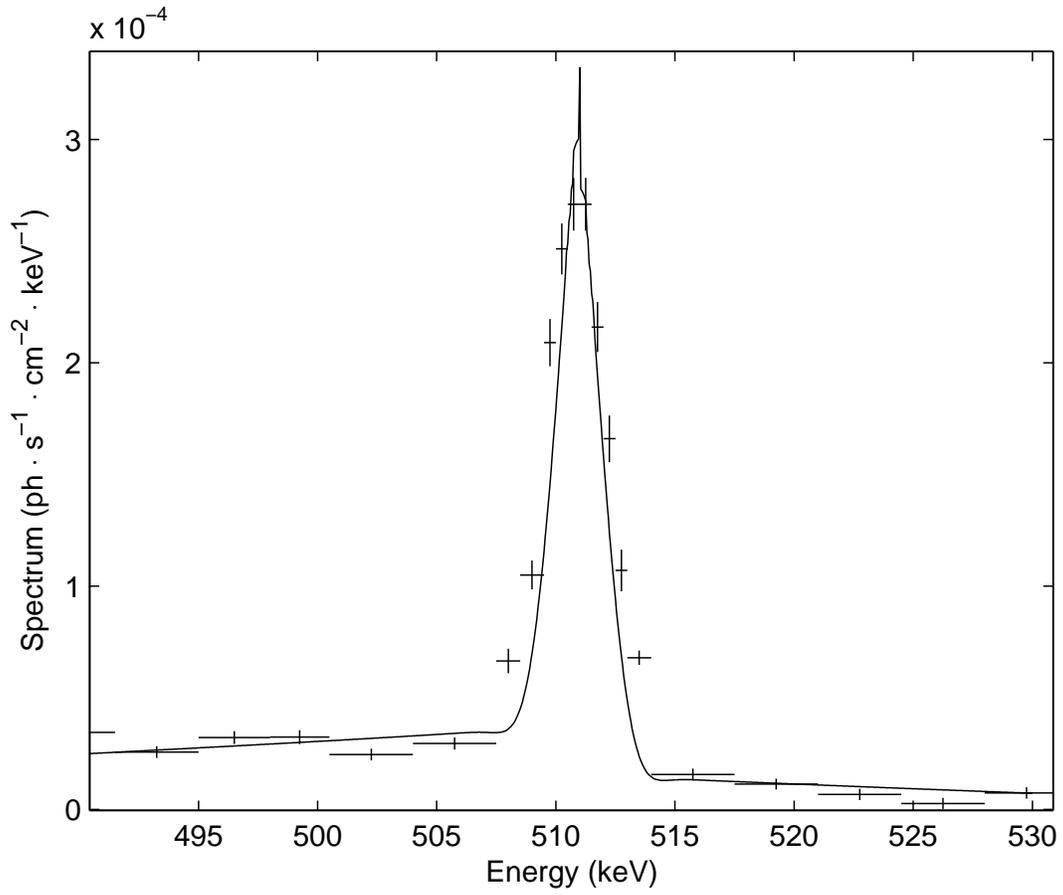} \caption{The total annihilation spectrum from a
plasma with a temperatures 2.5 eV, together with the INTEGRAL
data} \label{tflux}
\end{figure}

\newpage
\begin{figure}[ht]
\plotone{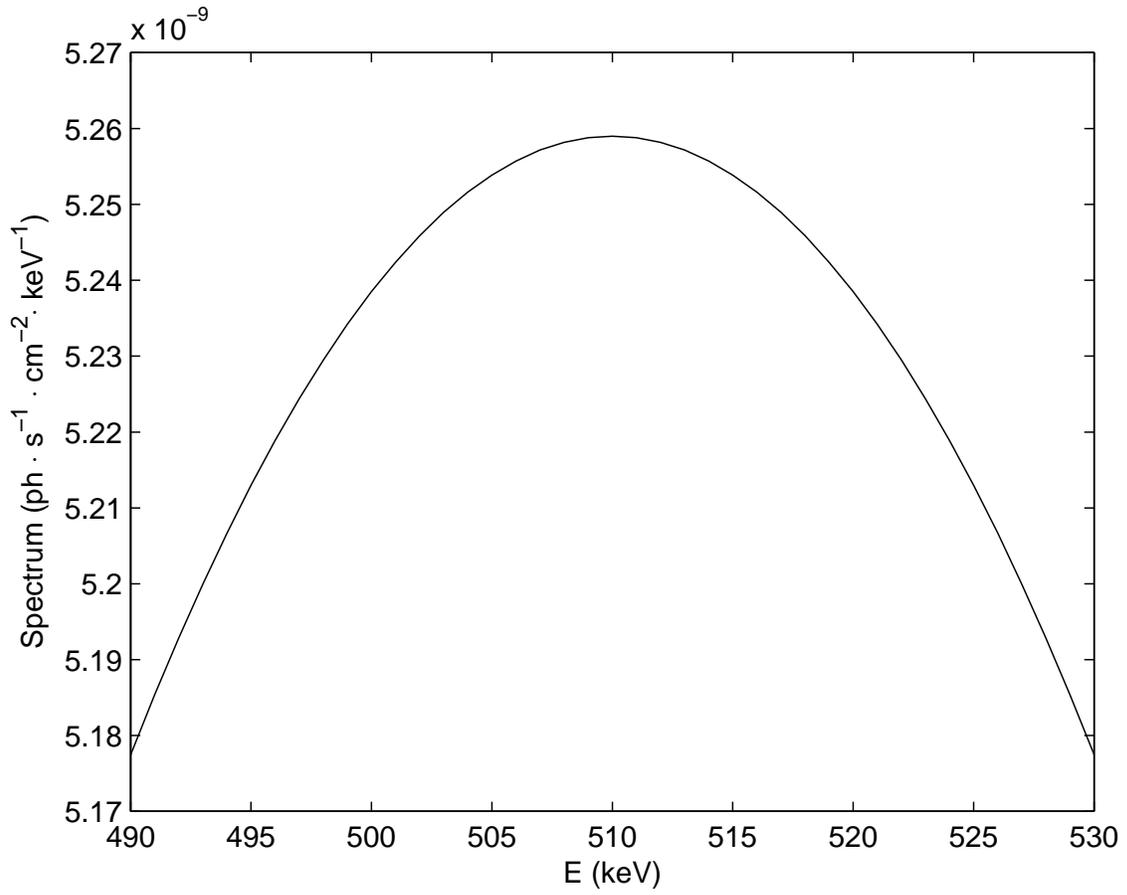}\caption{The annihilation flux from the hot plasma of the
central region of  radius $\sim 0.7$ kpc with a temperature of 100 eV, a
density of $3\times 10^{-3}$ cm$^{-3}$ and a filling factor of 30\%.}
 \label{an}
\end{figure}

\newpage
\begin{figure}[ht]
\plotone{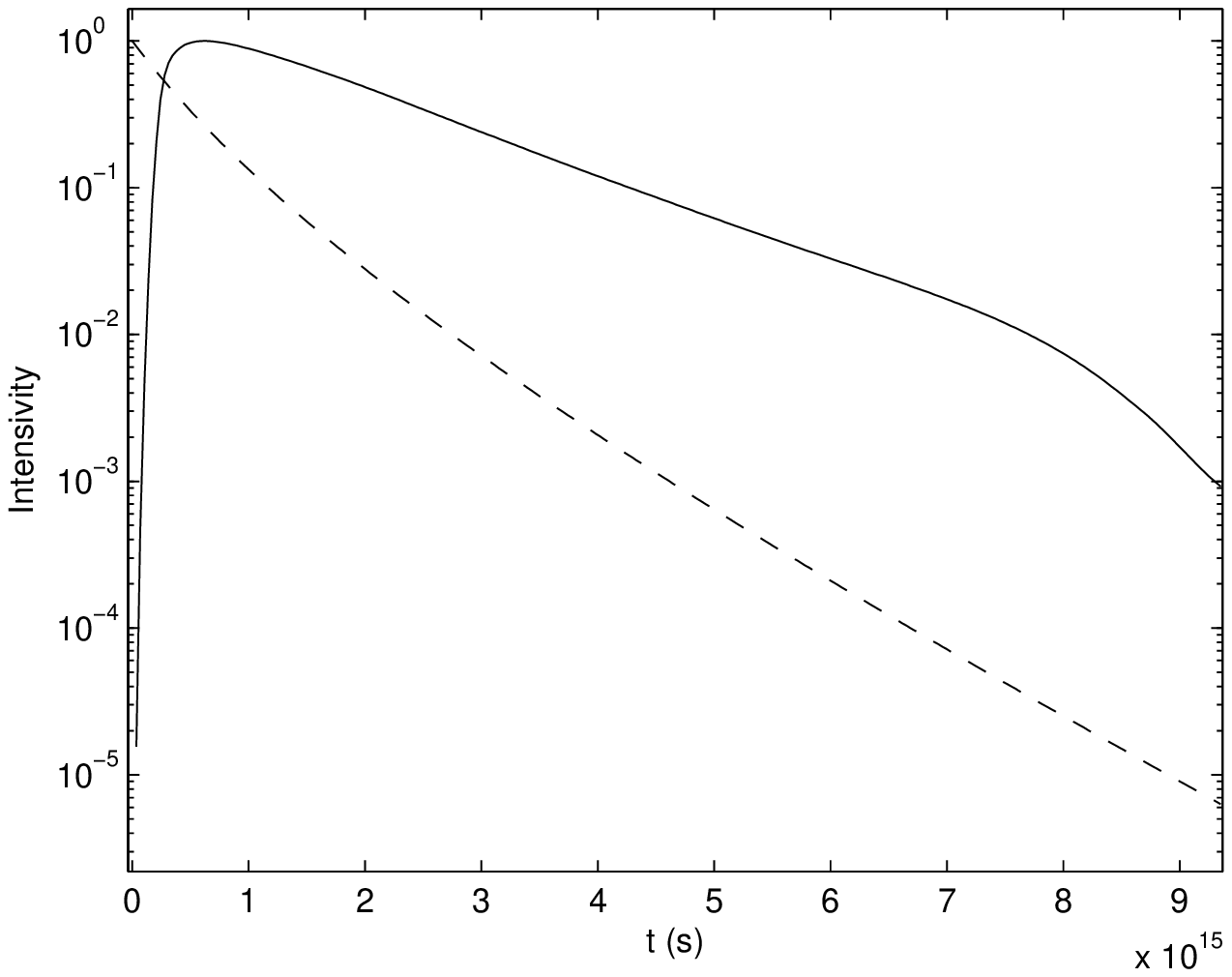} \caption{Time variation of the gamma-ray (dashed line) and
annihilation (solid line) fluxes for a gas density $n=1$ cm$^{-3}$ }
\label{2em}
\end{figure}
\newpage

\begin{figure}[ht]
\plotone{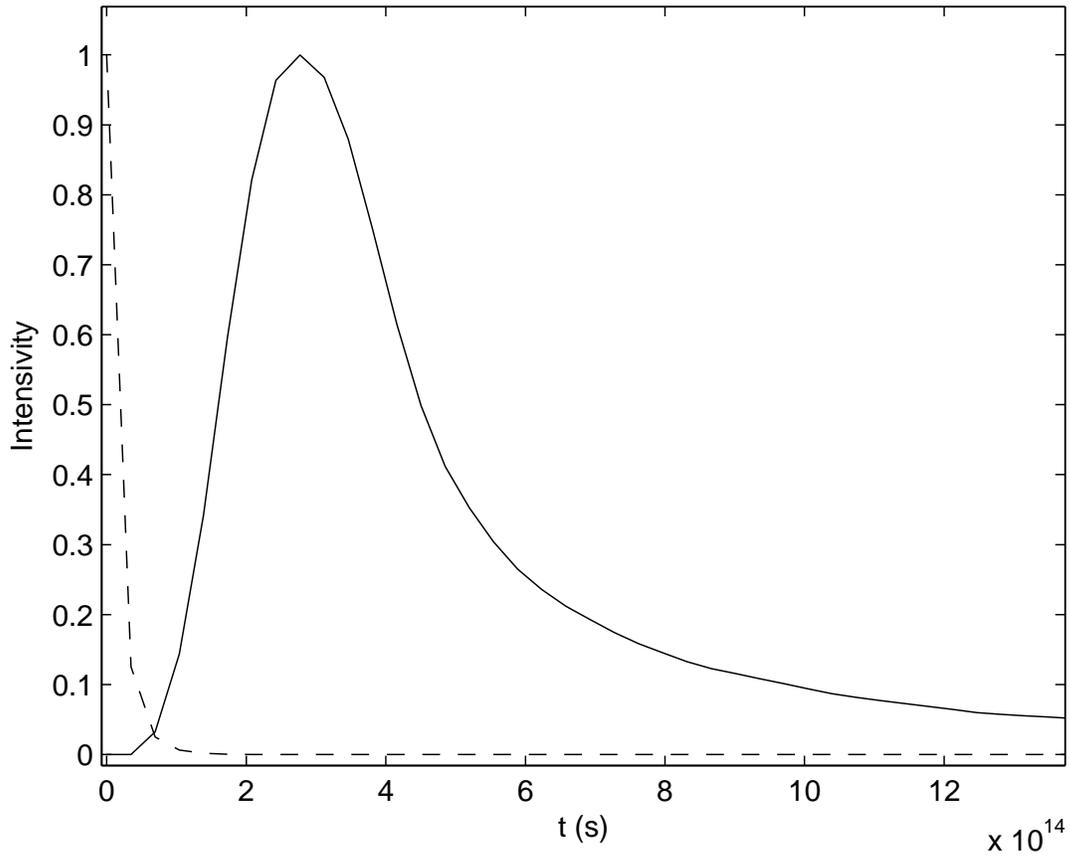} \caption{Time variation of the gamma-ray (dashed line)
and annihilation (solid line) fluxes  for a gas density $n=30$ cm$^{-3}$ for
protons} \label{var_30}
\end{figure}
\newpage
\begin{figure}[ht]
\plotone{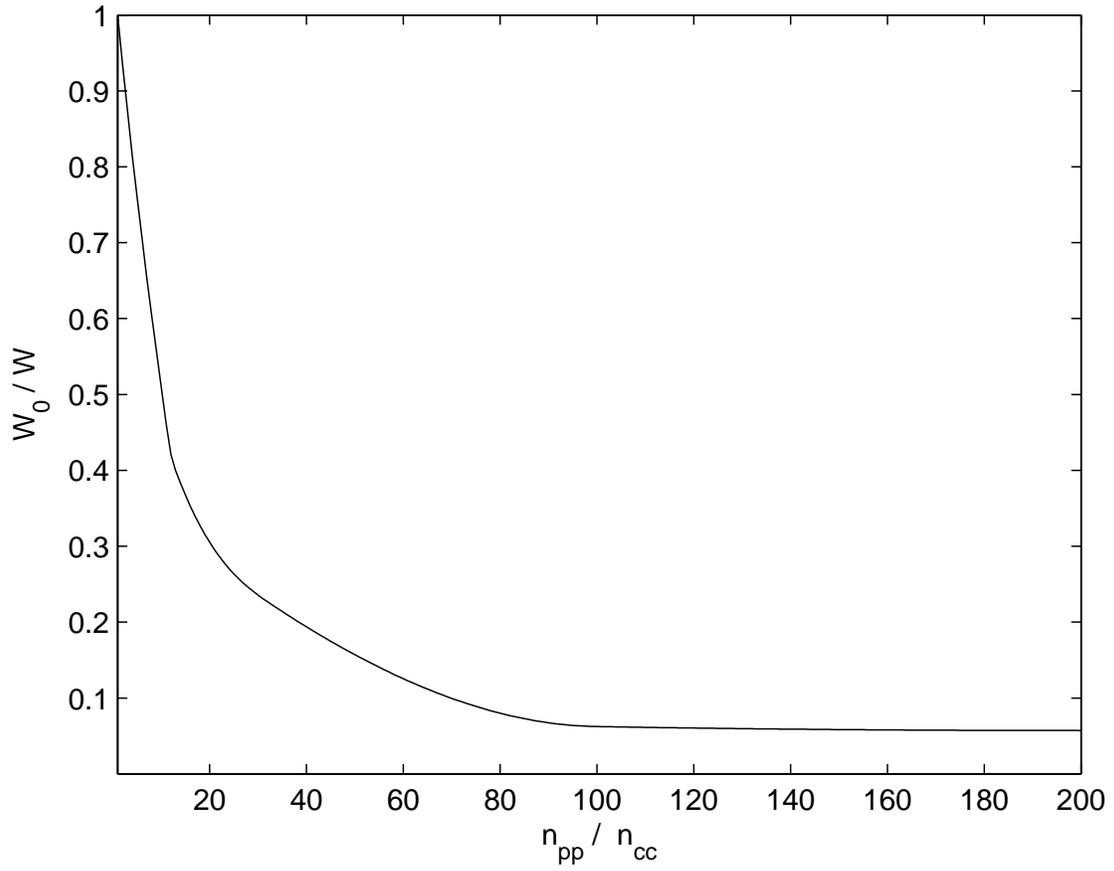} \caption{Variation of the proton energy output as a
function of the ratio $n_{pp}/n_{cc}$. The output energy $W_0$ corresponds to
$n_{pp}/n_{cc}=1$} \label{npp}
\end{figure}
\newpage
\begin{figure}[ht]
\plotone{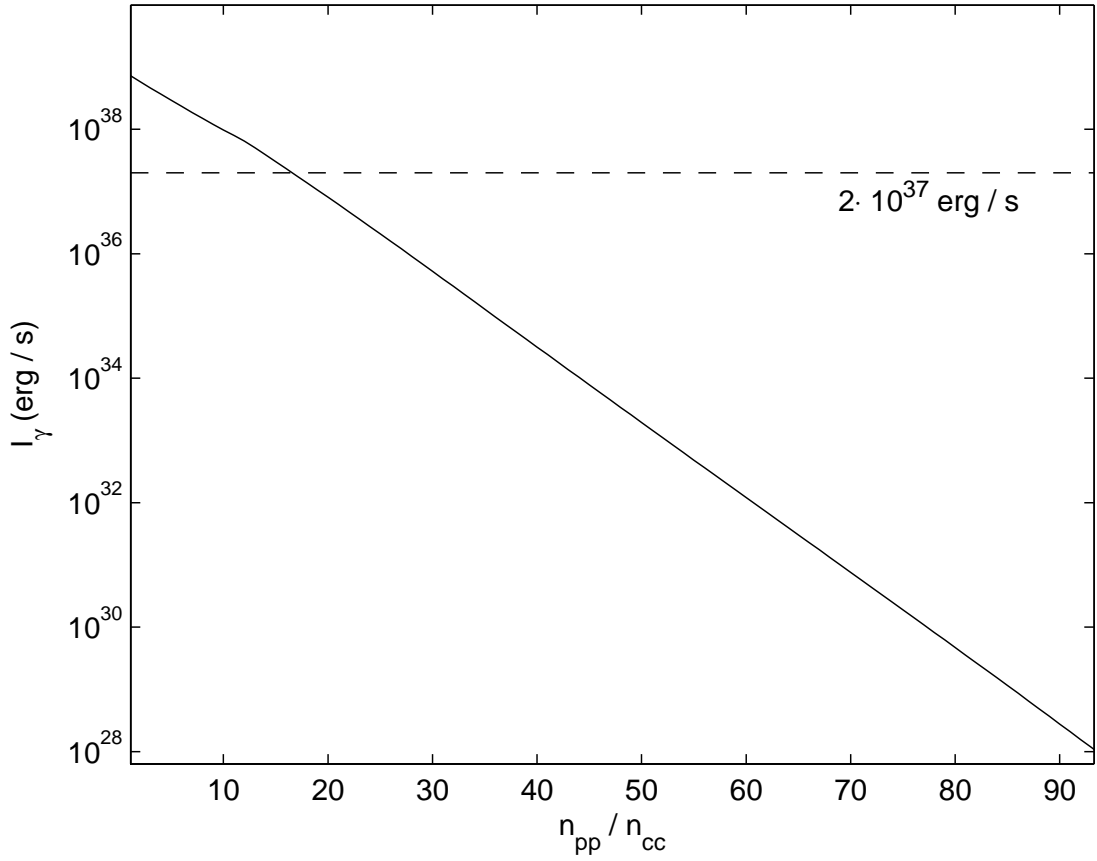} \caption{The expected gamma-ray flux at the moment when the
production of thermal positrons  reaches a value of $10^{43}$ e$^+$s$^{-1}$
as a function of the ratio $n_{pp}/n_{cc}$ } \label{g_npp}
\end{figure}
\newpage
\begin{figure}[ht]
\plotone{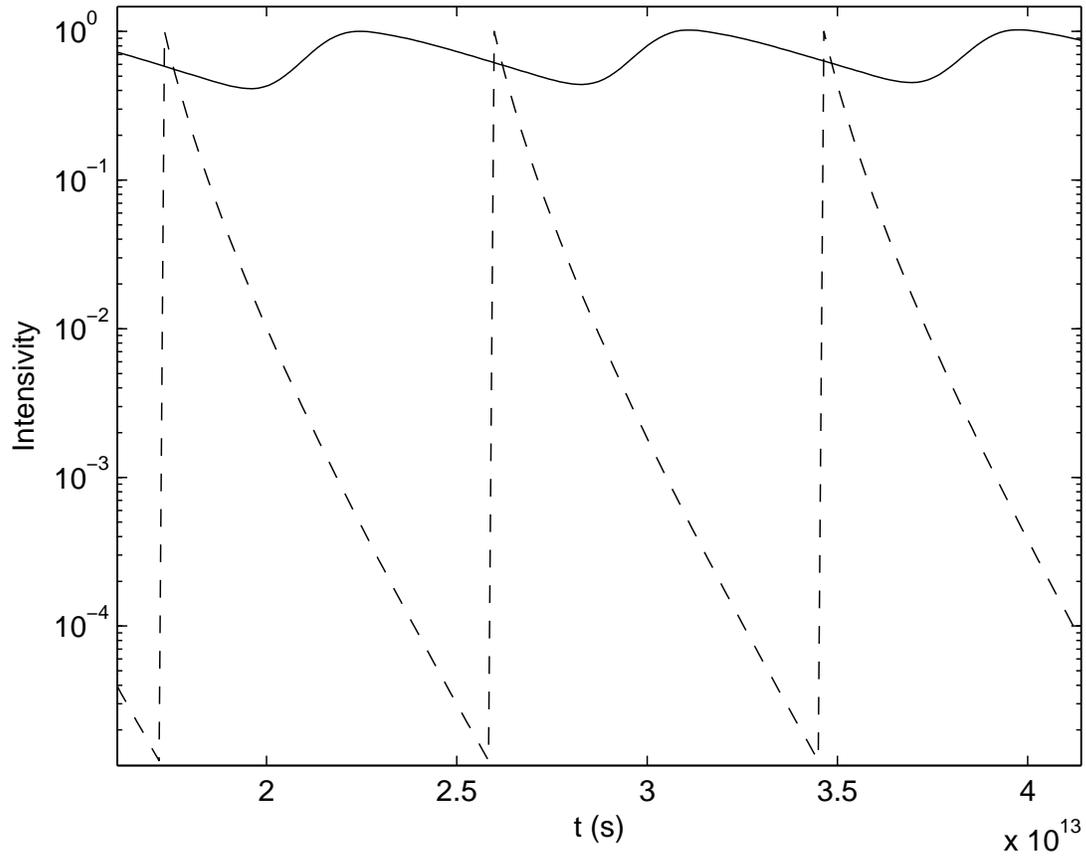} \caption{Gamma-ray and annihilation emission from several
successive eruption of protons by the black hole. The parameters are presented
in the text } \label{ncu3}
\end{figure}
\end{document}